\documentclass[a4paper,fleqn,usenatbib]{mnras}

\usepackage{graphicx} 
\usepackage{caption}
\usepackage{natbib}
\usepackage[export]{adjustbox}
\usepackage{subcaption}
\usepackage{multirow}
\usepackage{threeparttable}
\usepackage{rotating}
\usepackage{color}
\usepackage{alphalph}
\usepackage{footnote}
\usepackage{courier}
\usepackage{txfonts}

\usepackage[T1]{fontenc}
\usepackage{ae,aecompl}

\usepackage{graphicx}	
\usepackage{amssymb}	

\title[Observational Bias of GMCs]{Effects of Galactic Disc Inclination and Resolution on Observed 
GMC Properties and Larson's Scaling Relations}

\author[Pan et al.]{Hsi-An Pan$^1$, Yusuke Fujimoto$^1$, Elizabeth J. Tasker$^1$, Erik Rosolowsky$^2$,
  \newauthor 
   Dario Colombo$^{2,3}$, Samantha M. Benincasa$^4$, and James Wadsley$^4$ \\
  $^1$ Department of Physics, Faculty of Science, Hokkaido University, Kita 10 Nishi 8 Kita-ku, Sapporo 060-0810, Japan\\
  $^2$ Department of Physics, 4-181 CCIS, University of Alberta, Edmonton, AB T6G 2E1, Canada\\
  $^3$ Max Planck Institut f{\"u}r Radioastronomie, Auf dem H{\"u}gel 69, D-53121, Bonn, Germany \\
  $^4$ Department of Physics and Astronomy, McMaster University, Hamilton, ON, L8S 4M1, Canada
}

\date{Accepted XXX. Received YYY; in original form ZZZ}

\pubyear{2015}

\begin{document}
\label{firstpage}
\pagerange{\pageref{firstpage}--\pageref{lastpage}}
\maketitle

\begin{abstract}
With ALMA making it possible to resolve giant molecular clouds (GMCs) in other galaxies, it is becoming necessary to quantify the observational bias on measured GMC properties.
Using a hydrodynamical simulation of a  barred spiral galaxy,  we compared the physical properties of GMCs formed in position-position-position space (PPP) to the observational position-position-velocity space (PPV).
We assessed the effect of disc inclination: face-on (PPV$_{\mathrm{face}}$) and edge-on (PPV$_{\mathrm{edge}}$), and resolution: 1.5 pc versus 24 pc, on  GMC properties and the further implications of using  Larson's scaling relations for mass-radius and velocity dispersion-radius.
The low-resolution PPV data are generated by simulating ALMA Cycle 3 observations using the CASA package.
Results show that the median properties do not differ strongly between  PPP and   PPV$_{\mathrm{face}}$ under both resolutions, but PPV$_{\mathrm{edge}}$ clouds deviate from these two. The differences   become magnified when switching to the lower, but more realistic resolution. The discrepancy  can lead to opposite results for the  virial parameter's measure of gravitational binding, and therefore the dynamical state of the clouds.  
The power-law indices for the two Larson's scaling relations decrease going from PPP, PPV$_{\mathrm{edge}}$ to PPV$_{\mathrm{face}}$ and decrease from high to low resolutions. We conclude that the relations are not entirely driven by the underlying physical origin and therefore have to be used with caution when considering the environmental dependence,   dynamical state, and the extragalactic CO-to-H$_{2}$ conversion factor of GMCs.
\end{abstract}



\section{Introduction}
\label{SEC_intro}

As many observations show that star formation efficiency varies by $\sim$ 100 times from galaxy to galaxy, increasing attention has been devoted to the questions of whether giant molecular cloud (GMC) properties are universal and whether their  star formation ability depends on the large-scale galactic environment.
The Atacama Large Millimeter/submillimeter Array (ALMA) observations are starting to resolve a wide population of GMCs with the quality of that in the Milky Way. This gives a chance to reveal the whole picture of the relation between GMCs and star formation in various  environments. 
To interpret the measured GMC properties properly, it is necessary to understand any observational bias on the GMC properties.

It is still debated if the physical properties of    continuous structure in the  interstellar medium (ISM)   measured by spectral line observations (e.g., $^{12}$CO, HCN) represent the  intrinsic structures in three-dimensions \citep[e.g.,][]{Adl92,Pic00,Ost01,She08,She10,War12,Bea13,Pan15}.
Observations take  data from the galaxies projected on the sky plane. This provides  two spatial dimensions (RA and Dec.) and  a  velocity along the line of sight (LOS), known as the spectral data cube in Position-Position-Velocity space (PPV).
However, the non-spherical shapes of GMCs  change their appearances when  projected along different  LOS, which are determined by the  inclination and position angle of the host galactic disc. 
Moreover,  galaxies are distributed over a wide range  in distance in the Universe, causing the physical resolution of  observations to not always be the same.
Finite resolution will necessarily introduce  contributions from separated adjacent structures to the measured GMC properties.

These effects are  assessable by comparing  GMC properties in observations to simulations.
Simulations normally have data with three-dimensional position ($x$, $y$, $z$) and velocity ($v_{\mathrm{x}}$, $v_{\mathrm{y}}$, $v_{\mathrm{z}}$) coordinates, so called Position-Position-Position space (PPP), from which GMC properties can be directly calculated. 
In  previous work,  we compared the physical properties of the GMCs identified in PPP and those in PPV in the same simulated galaxy, assuming  ideal circumstances, i.e., face-on observation with high resolution (1.5 pc) and sensitivity (1 K per 1 km s$^{-1}$) \citep{Pan15}.
The results show that PPP and PPV can potentially identify the same objects with closely matched properties within a scattering in value of  a factor of 2.   Yet such high   resolution and sensitivity are very difficult to achieve in extragalactic observations even with ALMA.

In this work, we further evaluate the effects of galactic disc inclination and observed resolution. As in \cite{Pan15}, this is done by comparing GMCs  in PPP and   PPV.
This paper is organized as follows. 
Method and datasets are introduced in Section \ref{sec_method}.
The cloud identification methods and the derivation of physical properties are introduced in Section \ref{sec_def_gmc}. 
Section \ref{sec_result} presents the results of this work.
We summarize the key results of our analysis in Section \ref{sec_summary}.

\section{Method and Datasets}
\label{sec_method}

The simulated galaxies were modelled on the barred spiral (SABc) galaxy, M83, using observational data from the 2MASS $K$-band image to estimate the stellar potential. The simulations were run using the three dimensional adaptive mesh refinement (AMR) hydrodynamics code, ENZO \citep{Bry14}.  The high-resolution (1.5 pc) simulation is presented in \cite{Fuj14}, along with a full description of the run parameters. The gas radiatively cooled down to 300 K  but no star formation or feedback was included.
 Typical temperatures in the GMCs are about 10 K, an order of magnitude below our minimum radiative cooling temperature. However, our resolution is not sufficient to resolve the full turbulent structure of the gas and we also do not include pressure from magnetic fields. Including a temperature floor of 300\,K therefore imposes a minimum sound speed of 1.8 km s$^{-1}$ on the gas to crudely allow for these effects. The velocity dispersion within our GMCs is typically higher than this by about a factor of 2  -- 3, implying that this floor does not have a significant impact on the cloud properties.
Previous work that has compared runs with and without star formation and feedback physics suggest that GMC properties are not strongly affected by these additions. We have therefore not included a star formation model in this work, focusing on the properties of the gas \citep{Tas09,Tas11,Tas15}. 

Face-on and edge-on projection images of the high-resolution simulated galaxy are shown  in the bottom and top panels of Figure \ref{FIG_sim_galaxy}. 
To assess the resolution effect, a  low-resolution (24 pc) simulation was made by decreasing  the total levels of refinement. The remaining run parameters are the same as in the high-resolution simulation.

There are six datasets in total, summarized in Table \ref{TAB_suumary_data}. Data structure  of PPP, edge-on observation in PPV, and face-on observation in PPV are prepared. 
Each dataset is produced at two resolutions:  1.5 pc (with  1 km s$^{-1}$ for PPV velocity axis) and $\sim$ 24 pc  (with 2 km s$^{-1}$). 
The high- and low-resolution PPP data using the aforementioned two simulations are presented as PPP$\mathrm{^{H}}$ and PPP$\mathrm{^{L}}$,  respectively.

The LOS of PPV data is the $z$-axis for the face-on case, and $y$-axis for the edge-on case.  The chosen resolutions will return clouds with fully and barely resolved properties respectively if the simulated clouds have properties similar to the Galactic GMCs.    Velocity resolutions are chosen so that  the  smallest clouds in both spatial resolutions can span across 2 -- 3 velocity channels if following the Larson's  relation (cf.,\S\ref{secsec_larson}).

The high-resolution PPV data are created from the high-resolution (1.5 pc) simulation.
The pixel size matches to cell size at 1.5 pc.
When identifying GMCs in the PPV dataset, we assume the galaxy is observed in $^{12}$CO (1--0) (115.2 GHz), the most commonly used transition for GMC observations. 
Therefore, only cells with a density greater than 100 cm$^{-3}$  were included in the data, corresponding to the excitation density of $^{12}$CO (1--0). 
This is consistent with the cloud identification of the PPP clouds, which selected using  a continuous contour at a density of 100 cm$^{-3}$.
The cloud identification algorithm of PPV expects the data to be emission intensity, rather than the gas density followed by simulations. We use a  Galactic CO-to-H$_{2}$ conversion factor ($X_{\mathrm{CO}}$) of 2 $\times$ 10$^{20}$ cm$^{-2}$ (K km s$^{-1}$)$^{-1}$ \citep{Bol13} to convert between the two. Since we do not consider chemistry nor radiative transfer and the excitation of molecular lines, the conversion factor cancels when we derive the cloud mass, so its precise value does not affect our results.  This also means that we have  a more accurate mass measurement than real observations,  and this work is thus an assessment purely for the unavoidable effects of projection, resolution, sensitivity and cloud identification method in  observations, i.e., any  discrepancy between the datasets can be largely attributed to these effects.
The high-resolution face-on and edge-on PPV data are referred as PPV$\mathrm{_{face}^{H}}$ and PPV$\mathrm{_{edge}^{H}}$, respectively. 
Comparisons between PPP$\mathrm{^{H}}$ and PPV$\mathrm{_{face}^{H}}$  have been presented in \cite{Pan15}.

In observations, the  image is formed by convolving the intrinsic structures of the observed target   with a two-dimensional Gaussian ``beam''.  The convolution plays a critical role in determining the appearance of the GMCs when the  beam size is $\geq$ GMC size.
To reproduce the convolution  in the low resolution PPV, the Common Astronomy Software Applications (CASA) package \citep{Mcm07} is used to simulate  ALMA Cycle 3 (2015 October -- 2016 September)  $^{12}$CO (1--0) observations.  
The input sky model   is the noise-free PPV$\mathrm{_{edge}^{H}}$ and PPV$\mathrm{_{face}^{H}}$ in unit of flux corrected by the $X_{\mathrm{CO}}$, assuming a distance and coordinate of M83.
 Therefore, all of PPV datasets are created from the simulation with 1.5 pc resolution.  
CASA task \emph{simobserve} is used to construct the $uv$ visibilities for the specified antenna configuration, then  \emph{simanalyze} is used to Fourier transform the $uv$ visibilities into the image plane. Hexagonal  mosaic is adopted for  mapping. The observing time of 12-m array is about 1 minute on each mosaic pointing. 
 The baselines range from 14.7 to 538.9 meters.
Note that we only use the 12-m array,  and the effect of interferometric observation, e.g., short-spacing problem, is kept  as part of the comparison  because most of extragalactic observations did not have corresponding single dish observation to combine.
 Around 60\% of the total flux (estimated from the noise-free PPV$\mathrm{^{H}}$) are missed. The observation is only sensitive to structures $<$ 500 pc, but is large enough to detect the GMC-scale structures. 
The final resolution of the low resolution PPV (PPV$\mathrm{_{edge}^{L}}$ and PPV$\mathrm{_{face}^{L}}$) are $\sim$ 24 pc ($\sim$ 1.3$\arcsec$) and 2 km s$^{-1}$.
The RMS noise level ($\sigma_{\mathrm{RMS}}$)   is  $\sim$ 10 mJy ($\sim$ 0.55 K), leading to a typical mass sensitivity (1 $\sigma_{\mathrm{RMS}}$ in 1 channel) of 2 $\times$ 10$^{3}$ M$_{\sun}$.

\begin{figure}
		\includegraphics[width=0.48\textwidth]{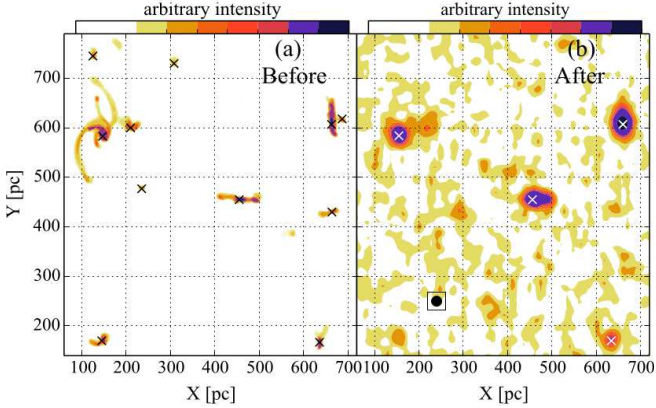}
	\caption{An example of how the ALMA processing affects the data.   Panel (a) and (b) show a $\sim$ 700 pc area of the face-on galaxy observed with high and low resolutions (PPV$\mathrm{_{face}^{H}}$ and PPV$\mathrm{_{face}^{L}}$), respectively.  Beam size ($\sim$ 24 pc) of low resolution is shown in panel (b) with solid black circle.  Centre of mass of GMCs are marked with  black or white crosses. It is clear that because of the low resolution or/and sensitivity in our setup, the small clouds in the high-resolution data  disappear in the low-resolution data. Moreover, clouds become more spherical in panel (b) due to the image convolution with the circular beam. }
	\label{FIG_Resolution}
\end{figure}

This observation setup is comparable to several ongoing ALMA projects on nearby galaxies.
In Figure \ref{FIG_Resolution}, we present a  side-by-side  showcase of how the ALMA processing affects the data.  
Panel (a) and (b) show  a $\sim$ 700 pc area seen in PPV$\mathrm{_{face}^{H}}$ and PPV$\mathrm{_{face}^{L}}$, respectively.
Comparison of two figures show that  small clouds in the high-resolution data disappear in the low-resolution data because of the lower resolution or/and sensitivity. Moreover, clouds become more spherical in panel (b) due to the image convolution with the circular beam.

\begin{table*}
\begin{minipage}{\textwidth}
\setlength{\tabcolsep}{-3pt}
\renewcommand{\arraystretch}{1}
\centering
\caption{Summary of the datasets in this work.}
\label{TAB_suumary_data}
\begin{tabular}{lcccccc}
\hline
                                  & \multicolumn{2}{c}{3D clouds}                                                                                                           & \multicolumn{2}{c}{edge-on observations}                                                                                                                           & \multicolumn{2}{c}{face-on observations}                                                                                                                           \\
\hline
                                  & PPP$^{\mathrm{H}}$                                                 & PPP$^{\mathrm{L}}$                                                 & PPV$\mathrm{_{edge}^{H}}$                                                    & PPV$\mathrm{_{edge}^{L}}$                                                           & PPV$\mathrm{_{face}^{H}}$                                                    & PPV$\mathrm{_{face}^{L}}$                                                           
                                  
                                  \\
\hline
Dataset final resolution {[}pc{]} & 1.5                                                                & 24                                                                 & 1.5\footnote{We use PPV$^{\mathrm{H}}$ to denote ``PPV$\mathrm{_{edge}^{H}}$ and PPV$\mathrm{_{face}^{H}}$''.}                                                                          & $\sim$ 24\footnote{We use PPV$^{\mathrm{L}}$ to denote ``PPV$\mathrm{_{edge}^{L}}$ and PPV$\mathrm{_{face}^{L}}$''}.                                                                               & 1.5                                                                          & $\sim$ 24                                                                               \\
Resolution of simulation {[}pc{]} & 1.5                                                                & 24                                                                 & 1.5                                                                          & 1.5                                                                                 & 1.5                                                                          & 1.5                                                                                 \\
PPV convolved by CASA             & $\dots$                                                            & $\dots$                                                            & N                                                                            & Y                                                                                   & N                                                                            & Y                                                                                   \\
Representation of resolution          & \begin{tabular}[c]{@{}c@{}}AMR level \\ of simulation\end{tabular} & \begin{tabular}[c]{@{}c@{}}AMR level \\ of simulation\end{tabular} & \begin{tabular}[c]{@{}c@{}}pixel size (MW- \\ and LG-like obs.\footnote{MW: Milky Way; LG: Local Group}
)\end{tabular} & \begin{tabular}[c]{@{}c@{}}beamsize from CASA\\  (extragalactic obs.)\end{tabular} & \begin{tabular}[c]{@{}c@{}}pixel size (MW- \\ and LG-like obs.)\end{tabular} & \begin{tabular}[c]{@{}c@{}}beamsize from CASA\\  (extragalactic obs.)\end{tabular} \\
Note                              & \cite{Pan15}\footnote{Detailed comparison of PPP$^{\mathrm{H}}$ and PPV$\mathrm{_{face}^{H}}$ are shown in \cite{Pan15}.}                                                            & $\dots$                                                            & $\dots$                                                                      & assume ALMA obs.\footnote{CASA can generate the visibilities  measured with  ALMA,  VLA, CARMA, SMA, and PdBI. ALMA is chosen for this work.}                                                                    & \cite{Pan15}     & assume ALMA obs. \\
\hline
\end{tabular}
\end{minipage}
\end{table*}

\section{Definition of Clouds in Simulation and Observation}
\label{sec_def_gmc}
The cloud identification methods are identical to \cite{Pan15}.
In both data structures, GMCs are identified as continuous structures of gas above a chosen density or flux threshold, and  multiple peaks are allowed within a GMC.
This method, which we refer to as the \emph{island} method in \cite{Pan15},  is the better choice for selecting similar objects between  simulation and observation data structures, compared to the \emph{decomposition} method which further segregates the peaks within an island into individual clouds.

 PPP works from the lowest density, drawing a contour at $n_{\rm HI, thresh} = 100$\,cm$^{-3}$ and defining all cells within a closed section as the cloud  \citep{Fuj14}. 
PPV clouds are identified using  CPROPS \citep{Ros06}.
The package was designed to identify continuous structures in the observed spectral data cube.
CPROPS begins by masking the emission with a high signal-to-noise ratio (S/N; 4 $\times$ RMS noise in this work), picking out the cloud locations at densities much higher than the background. It then extends this mask to the user defined lowest S/N (2 $\times$ RMS noise in this work), which outlines the observed cloud boundary. CPROPS then assumes that the real cloud boundary is larger than the observed cloud boundary, since the cloud outer regions are being obscured by the background noise. It therefore extrapolates  linearly from the observed boundary to a sensitivity of 0 K to form the real cloud boundary. 
 
Physical properties of the clouds are  derived once the cloud boundaries are set.
Derivations of physical properties in PPP and PPV  have been fully described in \cite{Fuj14} and \cite{Ros06}, respectively.
In this section, we provide a brief qualitative summary. 
These derivations are not identical between the two methods, since the  raw data  measure different quantities in different data structures. We do not  correct for this, but adopt the original calculations  as part of the comparison.

In both  data structures, cloud mass is a direct measured property from the sum of  cells or pixels enclosed within the  cloud boundary.
Radius and velocity dispersion of PPP clouds are calculated from three spatial and three velocity dimensions. 
For cloud radius, the average radius of the cloud is measured from its  projected area in the $x-y$, $x-z$, and $y-z$ planes.
The mass-weighted one-dimensional velocity dispersion of PPP clouds is computed from the average deviations between the gas velocity and the cloud bulk velocity in $x$, $y$, and $z$ direction.
PPV, however, must measure the mass(flux)-weighted projected radius at the plane perpendicular to the LOS (note that PPP does not consider any weighting in deriving radius) and the mass(flux)-weighted velocity dispersion along the  LOS using the second moments of the emission along the spatial and spectral axes.

The derived cloud properties, that depend on multiple cloud variables, are calculated from the three basic  properties above. Surface density, $\Sigma_{\mathrm{c}}$, is defined as the mass per unit area and is  simply calculated from the cloud mass and radius. Virial parameter, $\alpha_{\mathrm{vir}}$, measures the gravitational binding of a GMC, assuming a spherical profile and no magnetic support or pressure confinement. 
The classic derivation is defined as the ratio of virial mass ($M_{\mathrm{vir}}$) to cloud mass ($M_{\mathrm{c}}$) as $\alpha_{\mathrm{vir}}$  $\approx$ $M_{\mathrm{vir}}$/$M_{\mathrm{c}}$, where $M_{\mathrm{vir}}$ is calculated from the  cloud radius ($R_{\mathrm{c}}$) and velocity dispersion ($\sigma_{\mathrm{c}}$) as $M_{\mathrm{vir}}$ $\approx$ 1040 $R_{\mathrm{c}}$ $\sigma_{\mathrm{c}}^{2}$.
 $\alpha_{\mathrm{vir}}$ $>$ 2 indicates that the cloud
is gravitationally unbound while $\alpha_{\mathrm{vir}}$ $<$ 2 are  bound clouds \citep{Ber92}.

\section{Results}
\label{sec_result}

\subsection{Physical Properties of the GMCs in the High Resolution Analysis}
\label{secsec_phy_prop_gmc}

The probability distribution  function of  the high-resolution cloud properties are shown in the upper panel of Figure \ref{FIG_PDF_mass} -- \ref{FIG_PDF_alpha}.
Green-solid, red-dashed , and blue-dotted curves represent PPP$\mathrm{^{H}}$,  PPV$\mathrm{_{edge}^{H}}$ and PPV$\mathrm{_{face}^{H}}$, respectively. The coloured vertical lines indicate the median value of the properties for each dataset.

The median value   of the cloud mass  is consistent between the three datasets at $\sim$ 3 $\times$ 10$^{5}$ M$_{\sun}$, but  shapes of the  profiles    slightly differ. 
PPP$\mathrm{^{H}}$ and  PPV$\mathrm{_{face}^{H}}$ have almost identical mass profile.
\cite{Pan15} found that as high as 70\% of clouds have single counterpart in both data structures with a mass difference of $<$ 50\%.

The blending effect is  evident  for PPV$\mathrm{_{edge}^{H}}$ as demonstrated in  Figure \ref{FIG_Radial_PP4number}(a).
The figure shows   the galactocentric distribution of cloud numbers.
Line styles and colors are the same as in the 1D profiles.
Galactocentric distance ($R_{g}$) of PPP$\mathrm{^{H}}$ and PPV$\mathrm{_{face}^{H}}$ are calculated by their 3D and 2D position, respectively.
PPV$\mathrm{_{edge}^{H}}$, however, must use the kinematic distance.
If LOS velocities are known, clouds at a particular velocity can be assigned  a  position along the LOS, and thus a particular  $R_{g}$.  The method introduced by \cite{Yim11} is adopted for this. 
All three data structures detect two concentrations of clouds at radii $R_{g}$ $\approx$ 2 -- 3 kpc and $R_{g}$ $\approx$ 6 -- 7 kpc, corresponding to the radii of the galactic bar and the spiral arms. 
However,  the number of clouds in PPV$\mathrm{_{edge}^{H}}$ lies below both the PPP$\mathrm{^{H}}$ and  PPV$\mathrm{_{face}^{H}}$ cases by $\sim$ 2 times, indicating the  cloud blending.
Therefore, even though the  range and profile  of the cloud mass in PPV$\mathrm{_{edge}^{H}}$ do not deviate significantly from that of PPP$\mathrm{^{H}}$ and PPV$\mathrm{_{face}^{H}}$ overall,  this should not lead to the interpretation that the cloud mass between  the three data  sets are the same.

\begin{figure}
		\includegraphics[width=0.48\textwidth]{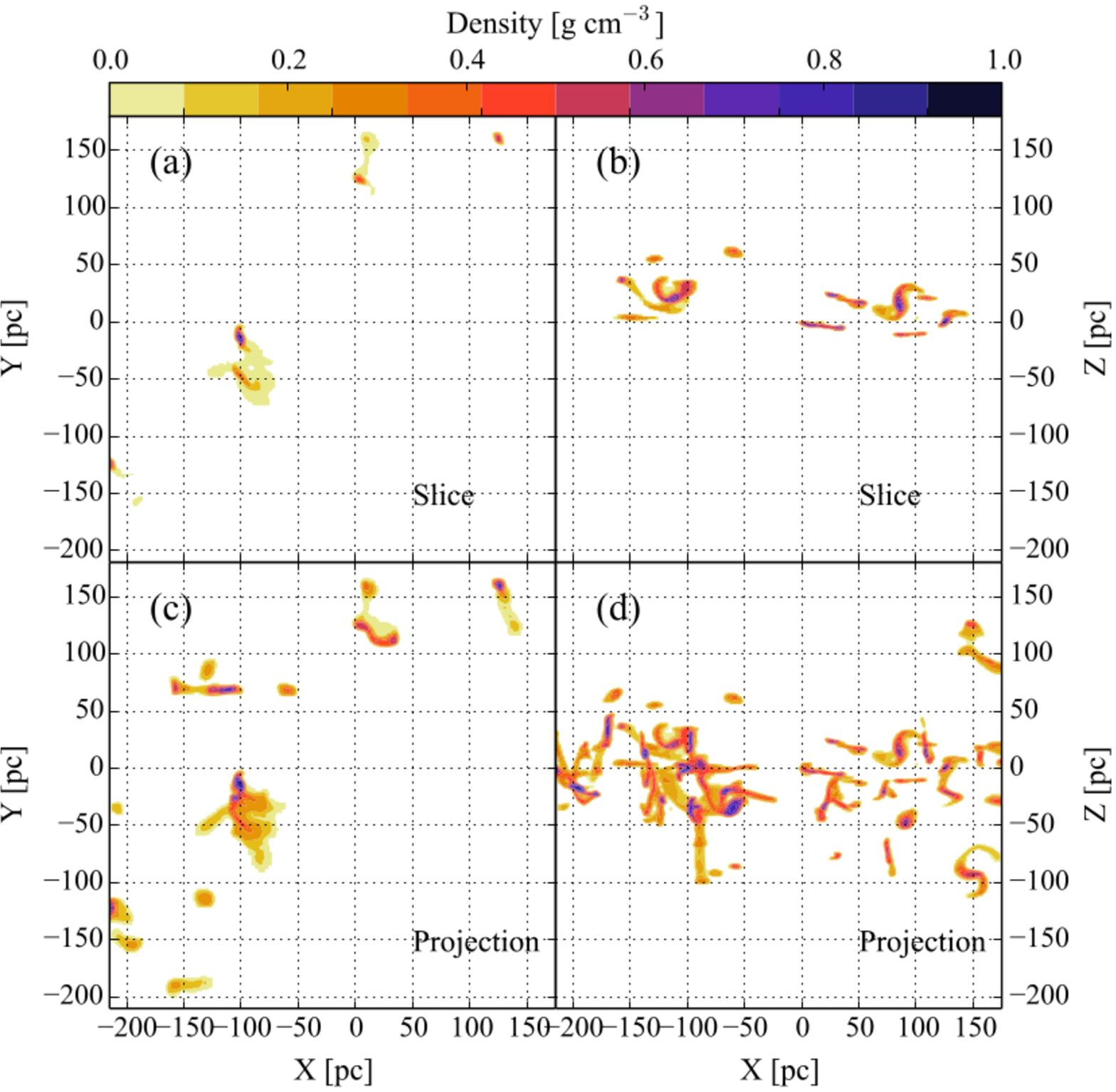}
	\caption{Example of GMCs seen from different viewing angle of $z$-axis (face-on, left) and $y$-axis (edge-on, right).  Note that y-axis ticks  of the edge-on cases (panel b and d) are shown at the right-hand side.  Only the cells with density $>$ $100$\,cm$^{-3}$  are plotted, therefore most of the coloured regions have been assigned to clouds.  Panel (a) and (b) present  the slice plots of face-on and edge-on views, respectively,  panel (c) and (d) show  projection plots of the the same orientations.The representative region is located at the spiral arm with galactocentric radius of $\sim$ 6 kpc. It can be seen from panel (b) that clouds are flat in the $z$-axis, and consequently, the slice and projection plots of PPV$\mathrm{_{face}^{H}}$ (panel (a) and (c)) show higher similarity and less crowding due to the small depth of the LOS, and vice versa for panel (b) and (d). }
	\label{FIG_ProjSliceCloud}
\end{figure}

The cloud radii identified in PPV$\mathrm{_{edge}^{H}}$ are smaller compared to the other two datasets (Figure \ref{FIG_PDF_rad}).
This is due to the  flat galactic disc (Figure \ref{FIG_sim_galaxy}), and is visualized in Figure \ref{FIG_ProjSliceCloud}.
Figure \ref{FIG_ProjSliceCloud}(a), (b) show the slice plots of a 400 pc patch of PPV$\mathrm{_{face}^{H}}$ and PPV$\mathrm{_{edge}^{H}}$, respectively, and the corresponding projection plots in panel (c) and (d).
Only the cells with density $>$ $100$\,cm$^{-3}$  are plotted, therefore most of the coloured regions have been assigned to clouds.
It can be seen from panel (b) that  clouds are flat in the $z$-axis, and consequently, the slice  and projection plots  of PPV$\mathrm{_{face}^{H}}$  (panel (a) and (c)) show higher similarity and less crowding due to the small depth of the LOS, and vice versa for panel (b) and (d).
The gas scaleheight of our simulated galaxy is about 80 -– 115 pc, which is similar to the initial value of 100 pc due to the lack of stellar feedback to inject energy; nevertheless, recent observations of edge-on spiral galaxies show that the scaleheights of molecular gas traced by $^{12}$CO (1-–0) are  mostly $<$ 150 pc \citep{Yim11,Yim14}, so this effect replicates true observations as well.
The velocity dispersion shows  very little difference between PPV$_{\mathrm{edge}}$ and PPP$^\mathrm{H}$/PPV$\mathrm{_{face}^{H}}$ (Figure \ref{FIG_PDF_vel}).
All of methods suggest a median value of $\sim$ 5 km s$^{-1}$.

In the analysis of this simulation  using the same cloud identification method, \cite{Fuj14} found that PPP$\mathrm{^{H}}$ clouds  fall into three  populations: the most common `Type A' clouds, with properties that corresponded to the average values measured in observations, the `Type B' massive giant molecular cloud associations that formed during repeated mergers of smaller clouds, and the transient `Type C' that were born in tidal tails and filaments.  
The three types  result in the bimodal distribution of cloud surface density  ($\Sigma_{\mathrm{c}}$) at $\sim$ 1000 M$_{\sun}$ pc$^{-2}$ (`Type A' and `Type B') and $\sim$ 100 $M_{\sun}$ pc$^{-2}$ (`Type C') as seen in the $\Sigma_{\mathrm{c}}$ profile of PPP$^\mathrm{H}$ (Figure \ref{FIG_PDF_sd}).

The clouds in the PPV$\mathrm{_{face}^{H}}$ data also split into two $\Sigma_{\mathrm{c}}$ regimes, as seen in Figure \ref{FIG_PDF_sd},  but the gap is not as sharp as in PPP$\mathrm{^{H}}$.
Fractions of each cloud type in the entire galaxy differ by less than 10\% between PPP$\mathrm{^{H}}$ and PPV$\mathrm{_{face}^{H}}$ \citep{Pan15}. 
Moreover, among the clouds that have a direct match in both datasets,  $\sim$ 80\% are categorized as same type between these two data structures.
On the contrary, the bimodal $\Sigma_{\mathrm{c}}$   is barely seen in PPV$\mathrm{_{edge}^{H}}$, presumably  due to the decrease of cloud radius   and cloud blending that increases $\Sigma_{\mathrm{c}}$.
 For the former effect, we would expect an increase of $\Sigma_{\mathrm{c}}$ for all GMCs, but it is not the case for the high-$\Sigma_{\mathrm{c}}$ population since the peak of its 1D profile is close to other two datasets. Therefore, the cloud blending  is likely in charge of  the missing bimodal-$\Sigma_{\mathrm{c}}$ in PPV$\mathrm{_{edge}^{H}}$, with the small transient populations  most susceptible to blending.

The median value of $\alpha_{\mathrm{vir}}$ $\approx$ 1 indicates that the majority of the clouds are bound in all three datasets.
The main difference is the extension of the profiles to lower values of $\alpha_{\mathrm{vir}}$ in two PPV$\mathrm{^{H}}$ sets. 
 The discrepancy arises because of the underestimation of cloud velocity dispersion in PPV$\mathrm{_{face}^{H}}$ and  cloud radius in PPV$\mathrm{_{edge}^{H}}$ when  the LOS goes along the short and long dimension of the clouds, respectively.
.

\subsection{Effect of Resolution}
\label{secsec_phy_prop_gmc_low_res}

In this section, we compare the low-resolution cloud properties between PPP$\mathrm{^{L}}$, PPV$\mathrm{^{L}}$. We emphasize again that the low-resolution PPP  and PPV clouds are identified from different simulations (\S\ref{SEC_intro} and Table \ref{TAB_suumary_data})  due to the different origin of ``resolution''. In simulations, resolution is set by the AMR level, which   locally refines the mesh to where they are needed.
On the other hand, resolution is determined by the size of the beam  that is used for convolving an object with finite (high) resolution in observation.

\subsubsection{PPP Clouds: Effect of Resolution (Cloud Blending)}
We start with the result of the PPP clouds, since it illustrates the  blending effect without any influence of the cataloging algorithms.  In other words, it represents the best possible case, though it is observationally infeasible.

The property  and galactocentric profiles are shown in the lower panels of Figure \ref{FIG_PDF_mass} -- \ref{FIG_PDF_alpha} and Figure \ref{FIG_Radial_PP4number}(b).
Line styles and colors are the same as in the high-resolution plots. 

The number of PPP$\mathrm{^{L}}$ clouds is smaller than that of PPP$\mathrm{^{H}}$.
This is a  result of  fewer refinement levels.
In spite of the different cloud numbers, both resolutions show  an accumulations of clouds at the bar ($R_{\mathrm{g}}$ $\approx$ 3 kpc) and spiral ($R_{\mathrm{g}}$ $\approx$ 6 -- 8 kpc) regions in Figure \ref{FIG_Radial_PP4number}(b).
This suggests that although the blending effect may alter cloud properties, simulations can potentially see the spatial distribution of clouds regardless of resolution.

For the cloud mass distribution in Figure \ref{FIG_PDF_mass}, the range of the profiles are comparable between PPP$\mathrm{^{H}}$ and PPP$\mathrm{^{L}}$, but PPP$\mathrm{^{L}}$ sees more massive clouds at $\sim$ 10$^{7}$ M$_{\sun}$ and  fewer small clouds at $<$   2 $\times$ 10$^{5}$ M$_{\sun}$ as a result of cloud blending. 
The median mass then increases by $\sim$ 5 times from PPP$\mathrm{^{H}}$ to PPP$\mathrm{^{L}}$.

The median cloud radius (Figure \ref{FIG_PDF_rad}) also increases to $\sim$ 40 pc in PPP$\mathrm{^{L}}$. 
This is because  the cells in the low resolution simulation  are larger, so clouds blend to become extended structures.
Resolution does not affect the velocity dispersion significantly in Figure \ref{FIG_PDF_rad},  where the range and profile of the distributions are similar between PPP$\mathrm{^{L}}$ and PPP$\mathrm{^{H}}$.

$\Sigma_{\mathrm{c}}$, the surface density, is significantly affected by  resolution.
The lower resolution simulation  blurs the distinction between the three cloud types as seen in Figure \ref{FIG_PDF_sd}.
PPP$\mathrm{^{L}}$ show a  uniform $\Sigma_{\mathrm{c}}$ at $\sim$ 200 M$_{\sun}$ pc$^{-2}$. 
This is in agreement with lower resolution studies performed by \cite{Tas09}.
The bimodal $\Sigma_{\mathrm{c}}$ no longer exists in PPP$\mathrm{^{L}}$,  suggesting that the clouds formed via different mechanisms (the three types cloud) cannot be differentiated at 24 pc resolution, even though the cloud properties are extracted directly from 3 dimensions.

The distribution of $\alpha_{\mathrm{vir}}$ suggests that the majority of PPP$\mathrm{^{L}}$  clouds are gravitationally bound as in PPP$\mathrm{^{H}}$. The profile of the distribution is also in good agreement between two resolutions.
Therefore, resolution may  not be a worrying issue  for the dynamical state of PPP clouds.

\subsubsection{PPV Clouds: The Combination of Resolution (Blending),  Sensitivity and Projection  Effects in Observer Space}

In addition to the resolution or blending effect, casting into observer space invokes  sensitivity and projection effects as well. Here we compare the results between  PPV$\mathrm{_{face}^{L}}$ and PPV$\mathrm{_{edge}^{L}}$, as well as their high-resolution counterparts and PPP clouds.

\begin{figure}
		\includegraphics[width=0.48\textwidth]{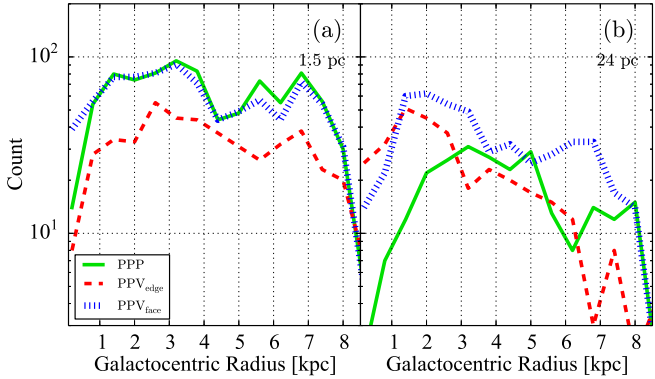}
		\caption{Galactocentric distribution of cloud numbers in  (a) high and  (b) low resolutions. Green-solid, red-dashed, and blue-dotted  curves represent the clouds from PPP, PPV$_{\mathrm{edge}}$, and PPV$_{\mathrm{face}}$, respectively. 
Galactocentric distance of PPP and PPV$_{\mathrm{face}}$ are calculated by 3- and 2-dimensional position, respectively, while 
 PPV$_{\mathrm{edge}}$ adopts   kinematic distance using the method introduced in \citet{Yim11}. In the high resolution,  the  cloud numbers are similar across all datasets, showing two concentrations at the radii of galactic bar and spiral arms. However, the cloud number of PPV$\mathrm{_{edge}^{H}}$ lies below PPP$^{\mathrm{H}}$ and PPV$\mathrm{_{face}^{H}}$ by roughly a factor of two, indicating the cloud blending. In the low resolution,  PPV$\mathrm{_{edge}^{L}}$ no longer represents the intrinsic distribution of clouds, but decline with radius, whereas PPV$\mathrm{_{face}^{L}}$ show a similar profile as PPP$^{\mathrm{L}}$ and their high-resolution counterparts, but with fewer clouds everywhere.
}
			\label{FIG_Radial_PP4number}
\end{figure}

As seen in the high resolution data in Figure \ref{FIG_Radial_PP4number}(a), the cloud number in the PPV$\mathrm{_{edge}^{L}}$ is smaller by a factor of $\sim$ 1.5 -- 2  compared to PPV$\mathrm{_{face}^{L}}$, except for the central 1 kpc area.
Moreover, it is notable that the galactocentric distribution of the PPV$\mathrm{_{edge}^{L}}$ dataset decreases with  radius, while PPV$\mathrm{_{face}^{L}}$    still observe two crowded regions around the galactic bar and spiral arms at $R_{\mathrm{g}}$ $\approx$ 2 -- 3 and 6 -- 8 kpc as seen in PPP$\mathrm{^{L}}$, but the cloud numbers are not the same.
This implies that with  low but realistic resolution,  edge-on observation no longer sees the real distribution of clouds, and the comparison between face-on observation (PPV$\mathrm{_{face}^{L}}$) and simulation (PPP$\mathrm{^{L}}$) should be carried out with great care as well.

\begin{figure*}
	\begin{subfigure}[]{0.32\textwidth}
	\centering
		\includegraphics[width=1.3\textwidth]{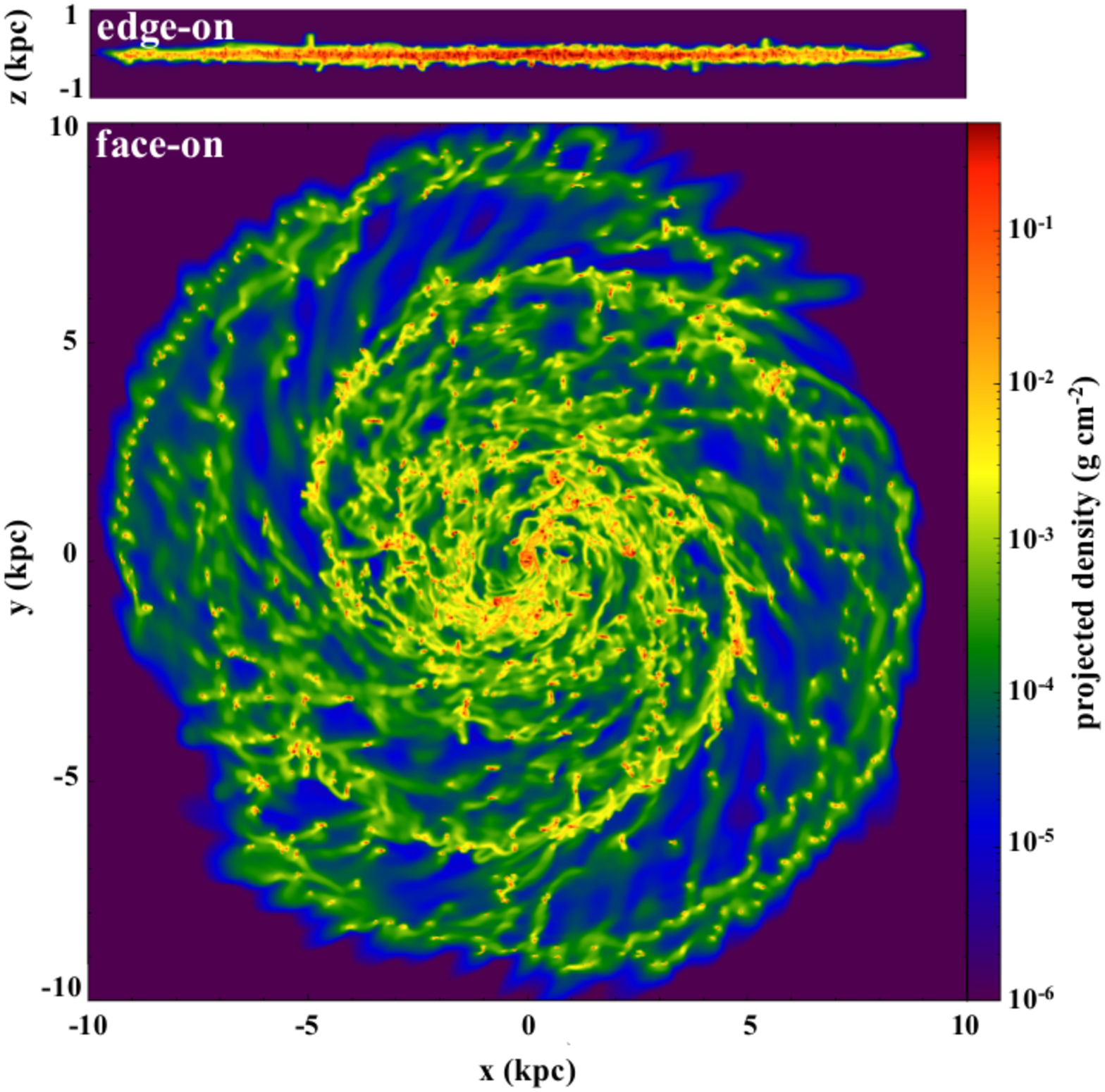}
		\caption{}\label{FIG_sim_galaxy}	
	\end{subfigure}
	\begin{subfigure}[]{0.32\textwidth}
		\includegraphics[width=1\textwidth]{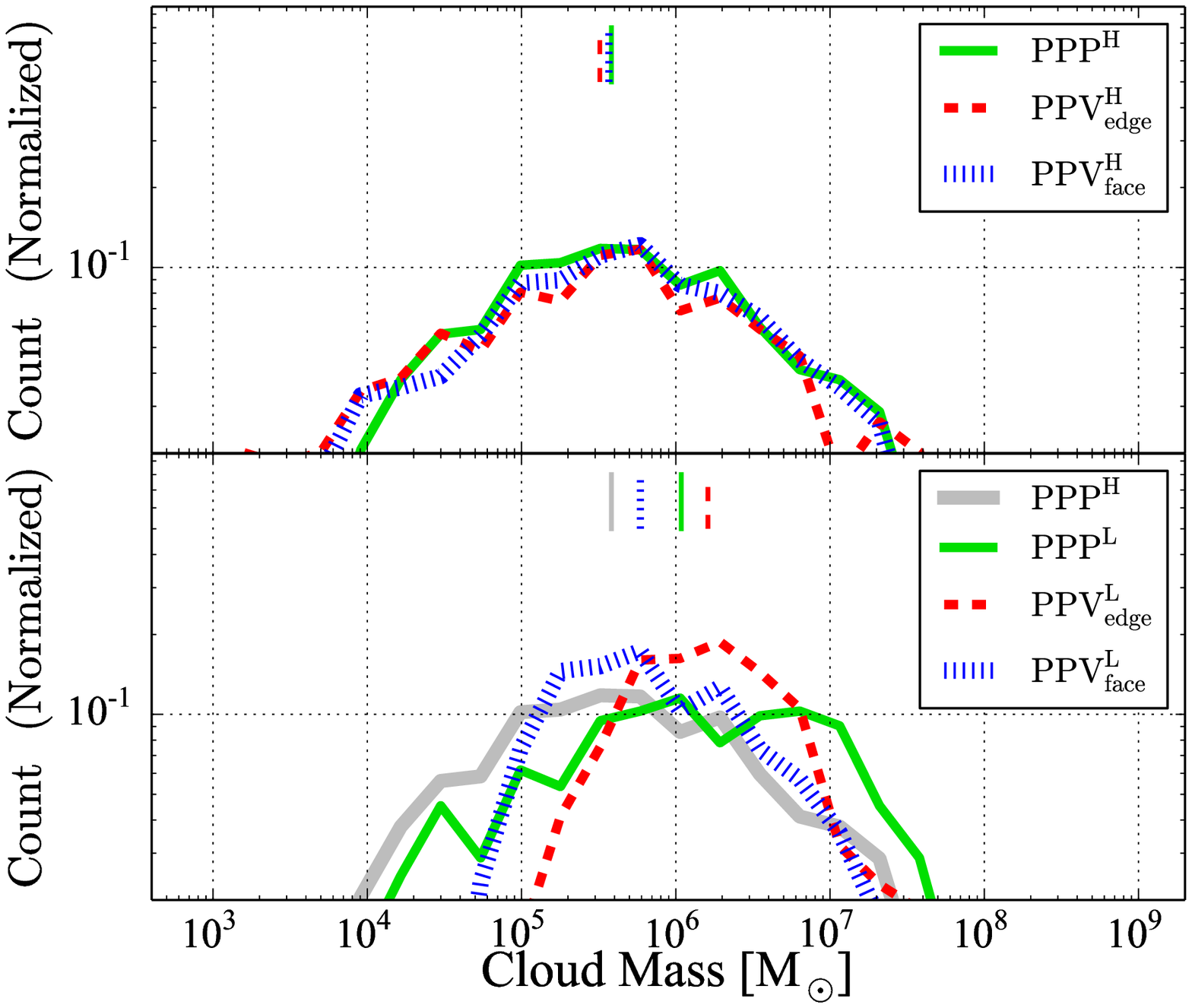} 
		\caption{}\label{FIG_PDF_mass}	
	\end{subfigure}
		\begin{subfigure}[]{0.32\textwidth}
		\includegraphics[width=1\textwidth]{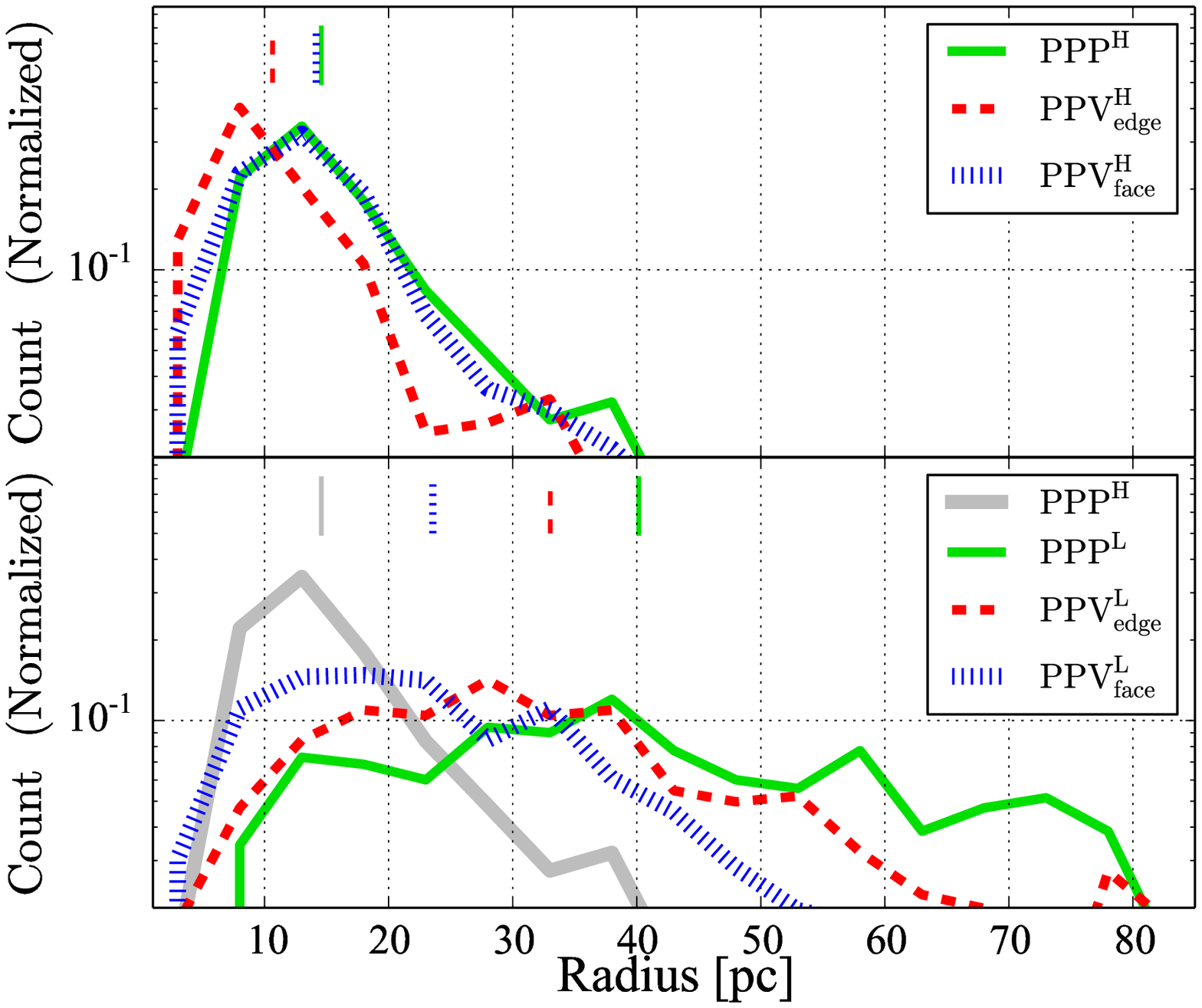} 
		\caption{}\label{FIG_PDF_rad}	
	\end{subfigure}
	\vspace{30pt}
			\begin{subfigure}[]{0.32\textwidth}
		\includegraphics[width=1\textwidth]{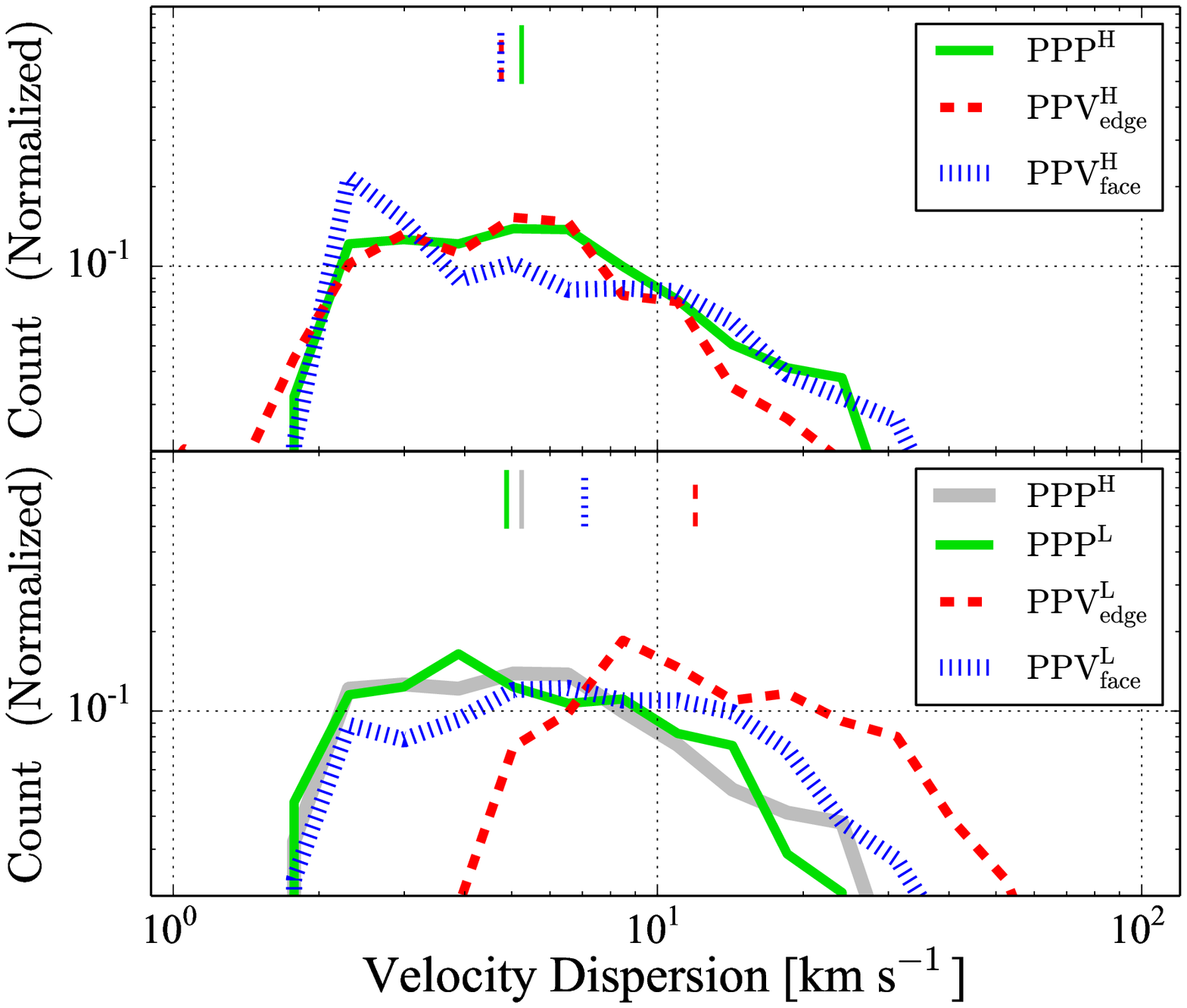} 
		\caption{}\label{FIG_PDF_vel}	
	\end{subfigure}
			\begin{subfigure}[]{0.32\textwidth}
		\includegraphics[width=1\textwidth]{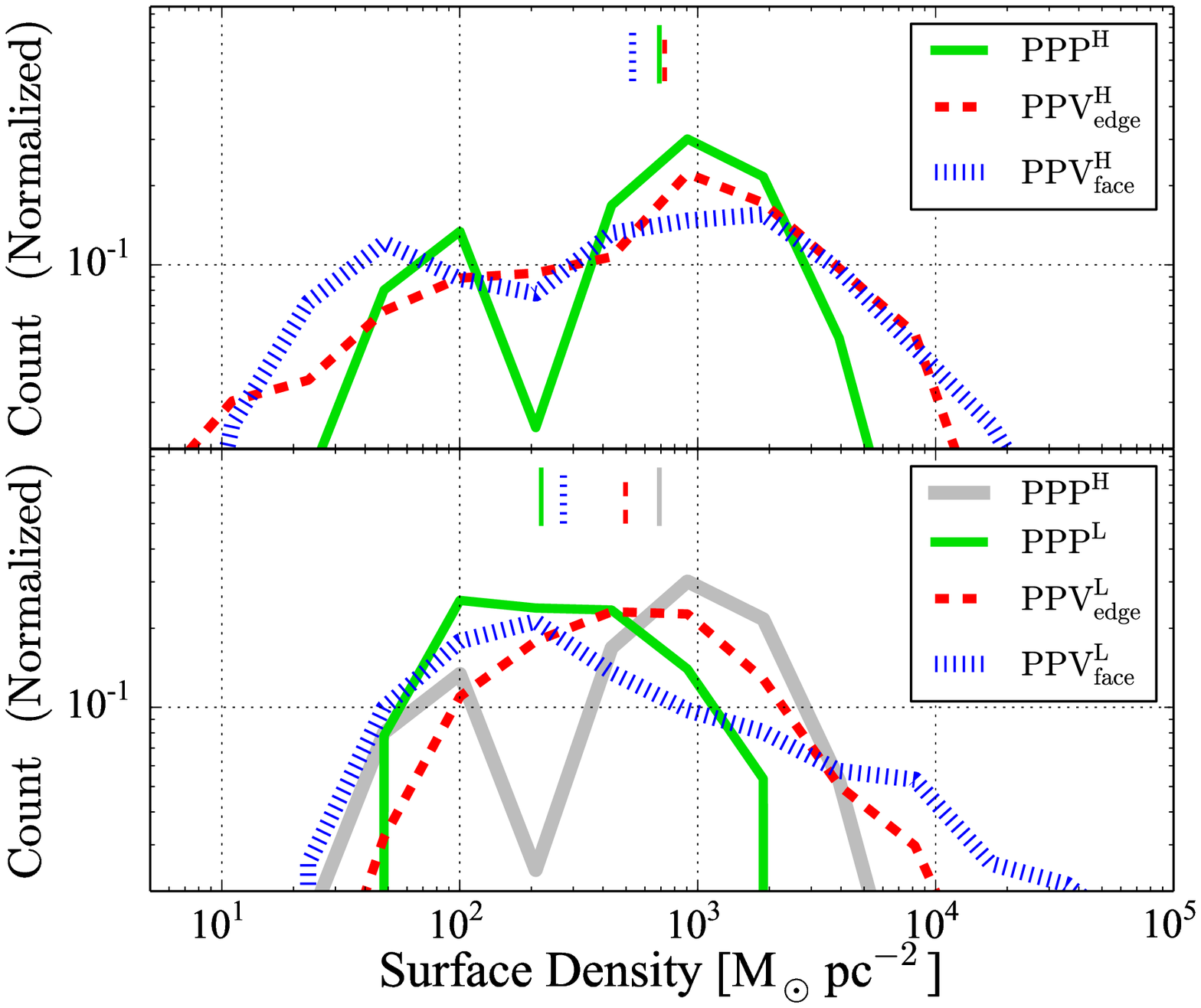} 
		\caption{}\label{FIG_PDF_sd}	
	\end{subfigure}
			\begin{subfigure}[]{0.32\textwidth}
		\includegraphics[width=1\textwidth]{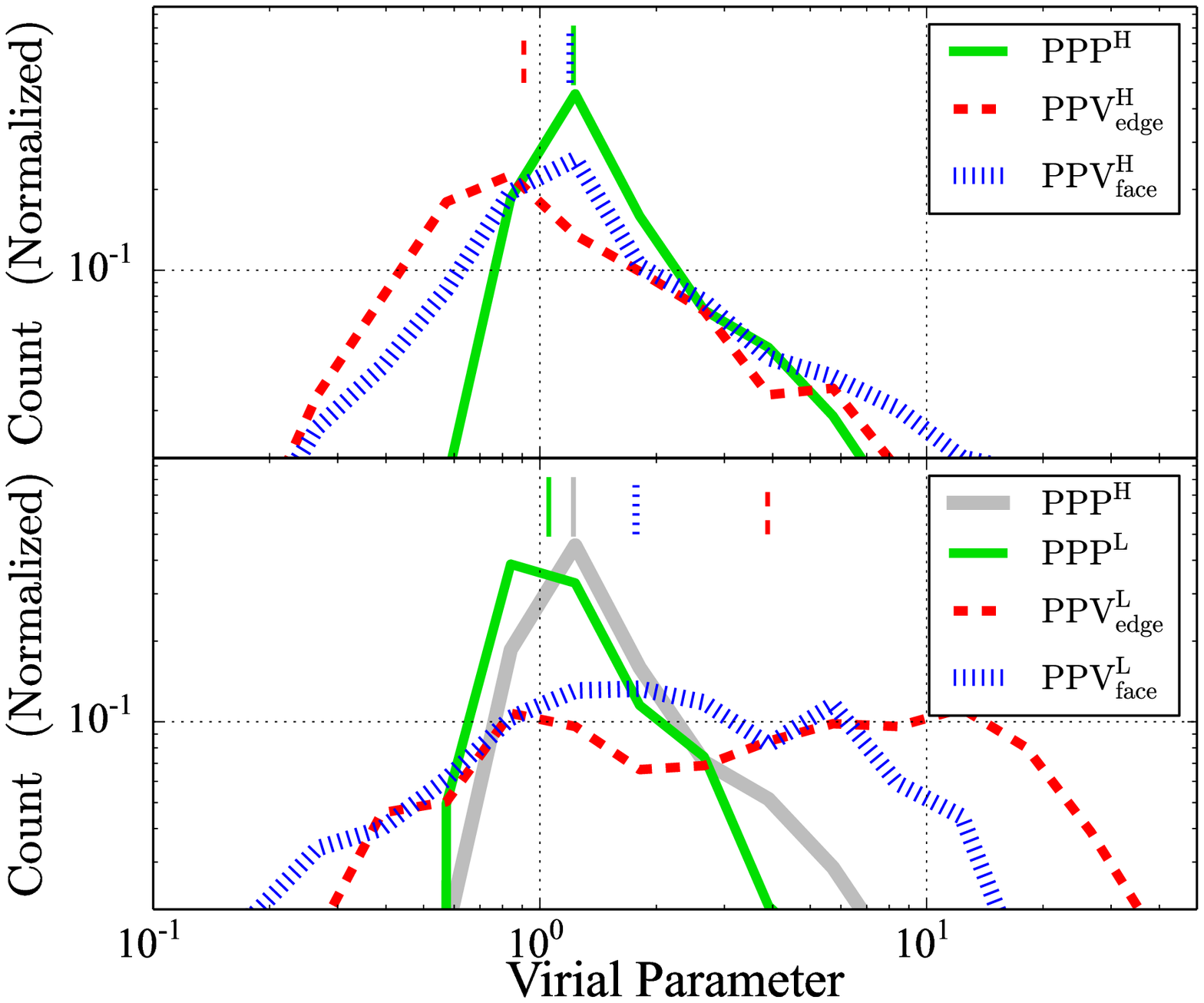} 
		\caption{}\label{FIG_PDF_alpha}	
	\end{subfigure}
	\caption{(a) Edge-on (top) and face-on (bottom) projection images of the simulated galaxy in high resolution. (b) Normalized distribution of cloud mass in high (top) and low (bottom) resolutions. Green-solid, red-dashed, and blue-dotted curves denote the distribution of PPP, PPV$_{\mathrm{edge}}$, and PPV$_{\mathrm{face}}$, respectively. In the bottom plots, we overlay the distribution of high-resolution PPP clouds (the green curves in the upper panels) to compare since it represents the ``real'' case. The coloured vertical lines denote the median value of each parameter. (c) Radius. (d) Velocity dispersion. (e) Surface density. (f) Virial parameter with the same segregation for both resolutions.}
	\label{FIG_PDFs}
\end{figure*}

The median values of  cloud mass in Figure \ref{FIG_PDF_mass} are consistent among  PPV$\mathrm{_{face}^{L}}$,  PPV$\mathrm{_{edge}^{L}}$, and  PPP$\mathrm{^{L}}$ within a factor of 2, but the  profile shapes are different.
The mass of the PPV$\mathrm{_{edge}^{L}}$ clouds are slightly larger than PPV$\mathrm{_{face}^{L}}$.
PPP$\mathrm{^{L}}$ shows a wider range in mass compared to PPV$\mathrm{^{L}}$ at both  ends.
At the higher end, the massive PPP$\mathrm{^{L}}$ clouds are the result of cloud blending as mentioned above. This can be seen in the cloud radius shown in Figure \ref{FIG_PDF_rad} as well (note that this is not exactly the same as convolving the high-resolution simulation to low resolution for making PPV$\mathrm{^{L}}$).
For the lower end,  our observational setups of PPV$\mathrm{^{L}}$ can only extract  properties from the clouds with mass $\geq$ 5 $\times$ 10$^{4}$ M$_{\sun}$, leading to the absence of small clouds.
 If  these data were observed with a perfect instrument to a lower noise level or higher sensitivity (but the same resolution of $\sim$ 24 pc), e.g., naively increase the integration time by 100 times, we would then  be able to see more small clouds, but most of them do not survive  the spatial and/or velocity deconvolution in CPROPS. Therefore, to obtain the small clouds (the  Type C clouds), both high resolution and high sensitivity are needed.

A caution that arises from this comparison is that the difference in the median mass among PPV$\mathrm{_{face}^{L}}$,  PPV$\mathrm{_{edge}^{L}}$, and  PPP$\mathrm{^{L}}$ are unlikely to be found in real observations. 
The differences are   comparable to  the uncertainty of $X_{\mathrm{CO}}$  derived from various methods \citep[][and reference therein]{Bol13}, i.e., projection effects can be obscured by adopting a different $X_{\mathrm{CO}}$.
We should bear in mind this essential point  when comparing GMC properties between  galaxies.

 The median cloud radius becomes $\sim$ 30 pc in  PPV$\mathrm{^{L}}$   as we progress to lower resolution. 
 The large median value compared to the high-resolution counterpart is mostly due to the  cloud blending, but note that small clouds also disappear due to insufficient sensitivity or resolution that pushes the median value toward larger end.
 The median radius of   PPV$\mathrm{^{L}}$ is slightly smaller than  PPV$\mathrm{^{H}}$.  
 In addition to the blending effect from the mesh refinement, the  large value of PPP$\mathrm{^{L}}$ is due in part to the derivation of cloud radius. We did not attempt to correct the resolution effect (e.g., deconvolution) in  PPP$\mathrm{^{L}}$, but keep the same derivation as our previous studies \citep{Fuj14,Pan15}, while CPROPS performs spatial deconvolution on cloud radius.
 The median velocity dispersion of PPV$\mathrm{_{edge}^{L}}$ clouds increases to $\sim$ 10 km s$^{-1}$, while it is 5 -- 6 km s$^{-1}$ in PPV$\mathrm{_{face}^{L}}$ (Figure \ref{FIG_PDF_vel}),  comparable to PPP$\mathrm{^{L}}$.
   It is the low resolution and blending effect cause the LOS to  pass through a considerably longer path within the PPV$\mathrm{_{edge}^{L}}$ clouds than for the other two datasets that increases the velocity dispersion significantly.

Both PPV$\mathrm{_{face}^{L}}$ and PPV$\mathrm{_{edge}^{L}}$  show a more uniform $\Sigma_{\mathrm{c}}$ (Figure \ref{FIG_PDF_sd}) as seen in PPP$\mathrm{^{L}}$.
The median values of $\Sigma_{\mathrm{c}}$  is $\sim$ 200 M$_{\sun}$ pc$^{-2}$ for  PPV$\mathrm{_{face}^{L}}$, which is again in good agreement with PPP$\mathrm{^{L}}$, while the value is about two times higher for PPV$\mathrm{_{edge}^{L}}$.
Both values are  very similar to that of the observed ``classic'' GMCs in the Milky Way and nearby galaxies.
Therefore we cannot rule out the possibility that the constancy of the observed $\Sigma_{\mathrm{c}}$ is due to low resolution.
Resolution has a smaller effect on $\alpha_{\mathrm{vir}}$ for  PPV$\mathrm{_{face}^{L}}$ (Figure \ref{FIG_PDF_alpha}). 
The majority of clouds are gravitaionally bound with median $\alpha_{\mathrm{vir}}$ of 1 -- 2 as suggested by PPP$\mathrm{^{L}}$, but note that the range  is  wider in PPV$\mathrm{_{face}^{L}}$.
In contrast, median $\alpha_{\mathrm{vir}}$ of PPV$\mathrm{_{edge}^{L}}$ increases to $\sim$ 4 mostly because  velocity dispersion is squared to calculate $\alpha_{\mathrm{vir}}$, leading to an opposite implication that overall the clouds are not bound.

\subsection{The Larson's Scaling Relations}
\label{secsec_larson}

The physical properties of GMCs are believed to regulate star formation. This was first observed by \cite{Lar81} using the  Galactic GMCs. He found that  GMCs  are characterized by $M_{\mathrm{c}}$-$R_{\mathrm{c}}$ and $\sigma_{\mathrm{c}}$-$R_{\mathrm{c}}$ scaling relations as:
\begin{equation}
 M_{\mathrm{c}}\propto R\mathrm{_{c}}^{a}, \:  \sigma_{\mathrm{c}}\propto R\mathrm{_{c}}^{b}.
\label{EQ_larsons}
\end{equation}
The relations and the power-law index $a$ and $b$ have been interpreted as evidence that GMCs are supported by  internal turbulence,  have constant $\Sigma_{\mathrm{c}}$ and  are gravitationally bound.
To date, Larson's scaling relations have been tested across different galaxy environments and  used to interpret whether GMC properties are universal \citep{Oka98,Hey09,Hug10,Reb12,Col14,Reb15,Swi15,Uto15,Ler15}. However,  no solid conclusion has been reached yet.
Theoretical studies also attempt to explain the Larson's relations using numerical simulations \citep{Vaz97,Ost01,Oss02,Bal11,Kri13,Fuj14,War15}.
They found that the constant $\Sigma_{\mathrm{c}}$ of GMCs may be the result of  saturation of the emission, an optically thick effect, or the limited dynamical range in density available in which the particular tracer can be excited. \citep{Keg89,Oss01,Bal02,Bal06}.
In addition, \cite{She10} show that   the power-law indices of the scaling relations vary between PPP and PPV clumps %
within a simulated GMC. 
The discrepancy arises from the   projection effect and the  difficulty in measuring  properties of non-spherical clumps  from observations.
If this is really the case, then the effects may be more significant for GMCs since they are presumably more structured with multiple clumps, cores and envelopes than the small interior clumps.

Figure \ref{FIG_larsons_high} and \ref{FIG_larsons_low} show the Larson's relations for the high- and low-resolution clouds respectively. From the top to bottom rows, we present the relations for $M_{\mathrm{c}}$-$R_{\mathrm{c}}$, $\sigma_{\mathrm{c}}$-$R_{\mathrm{c}}$, and $\alpha_{\mathrm{vir}}$-$R_{\mathrm{c}}$. The left to right columns present the results of  PPP,  PPV$_{\mathrm{edge}}$, and  PPV$_{\mathrm{face}}$, respectively. 
 The fit is performed using the \texttt{POLYFIT} function of  \texttt{Python}'s \texttt{NumPy} package. \texttt{POLYFIT}  can be used to fit a polynomial of specified order to data using a least-squares approach. The 1-$\sigma$ uncertainty is estimated from the covariance matrix of the fit.
The power-law index and the uncertainty of the first two relations are shown in  each panels. The dotted lines in $M_{\mathrm{c}}$-$R_{\mathrm{c}}$ relations denote $\Sigma_{\mathrm{c}}$ $=$ 50 (lower), 230, 1000, and 5000 (upper) M$_{\sun}$ pc$^{-2}$, respectively.

\subsubsection{Mass--Radius Relation}

 In all cases,  $M_{\mathrm{c}}$-$R_{\mathrm{c}}$ relation shows a strong correlation. 
In the high-resolution clouds, the best-fit line gives a power-law index  $a$ $\approx$ 3.6 for PPP$\mathrm{^{H}}$.
This  is a  result of the bimodal-$\Sigma_{\mathrm{c}}$, for which both  sequences have gradient of 3.
$a$ $\approx$  3 for 3D clouds indicates that the GMCs have constant   volume densities.
The same index is found for the PPP clumps of \cite{She10}.
For PPV$\mathrm{_{edge}^{H}}$ and PPV$\mathrm{_{face}^{H}}$, the best-fit indices with all clouds are $a$ $\approx$ 3.0 and 2.5, respectively.

Observations of the edge-on Milky Way normally found a value of 2.0 $<$ $a$ $<$ 3.0 \citep[e.g.,][]{Sol87,Sim01,Rom10}, our value is located in the higher end.  Face-on  observations of nearby galaxies have not achieved such high resolution.
Observations of the relatively face-on (barred) spiral galaxies LMC and M33  yielded an index $a$ of 2.23 $\pm$ 0.08 and 1.89 $\pm$ 0.16 with an observed resolution lower than our PPV$\mathrm{_{face}^{H}}$ data but higher than PPV$\mathrm{_{face}^{L}}$ \citep{Ros03,Won11}. 
The index of PPV$\mathrm{_{face}^{H}}$ is comparible to the simulated PPV clumps of \cite{She10}, where they also took a projection along the  $z$-axis.

The discrepancy between  PPV$\mathrm{_{edge}^{H}}$ and PPV$\mathrm{_{face}^{H}}$ originates from the different slopes of their  low-$\Sigma_{\mathrm{c}}$ clouds ($<$ 230 M$_{\sun}$ pc$^{-2}$)  that form in tidal tails and filaments, with  $a$ $=$ 3.19 $\pm$ 0.13 for PPV$\mathrm{_{edge}^{H}}$ and $a$ $=$ 2.44 $\pm$ 0.06 for PPV$\mathrm{_{face}^{H}}$. 
This is  because it is more difficult to fully identify the cloud boundary of small clouds in PPV space \citep{She10,Pan15}.
Moreover, some of the  low-mass PPV$\mathrm{^{H}}$ clouds are part of larger PPP$\mathrm{^{H}}$ clouds.
PPP$\mathrm{^{H}}$ cloud can be  split due to internal motions or the  projected density of substructures  below the noise level.
Finally, the  steep $a$ of the low-$\Sigma_{\mathrm{c}}$ PPV$\mathrm{_{edge}^{H}}$ clouds is also attributed to the rapid growth in cloud mass by the  blending and projection effect with respect to the confined cloud radius due to the flat disc.

For  the high-$\Sigma_{\mathrm{c}}$  clouds ($>$ 230 M$_{\sun}$ pc$^{-2}$), PPV$\mathrm{_{edge}^{H}}$ and PPV$\mathrm{_{face}^{H}}$ show similar slopes, with $a$ $=$ 2.00 $\pm$ 0.07 and $a$ $=$ 2.15 $\pm$ 0.10, respectively, implying that structures with larger mass are more likely to have  constant column densities.
However, this has to be considered with caution because there are  a few  factors that might have caused the uniform $\Sigma_{\mathrm{c}}$.
 The mass (flux) contrast between the core(s) and envelope of a PPV$\mathrm{_{edge}^{H}}$ cloud may be reduced  when the extended envelope are projected along a specific LOS that increases the observed $\Sigma_{\mathrm{c}}$, leading to a relatively uniform $\Sigma_{\mathrm{c}}$ on the surface of the cloud.
The aforementioned  mass weighting  causes PPV$\mathrm{_{face}^{H}}$ cloud radii to be dominated by the high-density (mass) cores, therefore the $M_{\mathrm{c}}$-$R_{\mathrm{c}}$ relation is dominated by particular regions \citep{Pan15}.
This effect  would be reduced to some extent in real observations since the dense regions are consumed to form stars when the density is sufficiently high. This means that the difference between $a$ of PPV$\mathrm{_{edge}^{H}}$ and PPV$\mathrm{_{face}^{H}}$ would enlarge because the relation of PPV$\mathrm{_{face}^{H}}$ would become shallower.
Taking account of these,  we cannot  conclude that the column density of  the high-$\Sigma_{\mathrm{c}}$ clouds  is equivalent everywhere  in these synthetic observations.

Moving to the low-resolution clouds in Figure \ref{FIG_larsons_low},   power-law indices $a$ decrease strongly when the  resolution is lowered. 
This is true for all data structures.
The power-law index of PPP$\mathrm{^{L}}$, PPV$\mathrm{_{edge}^{L}}$, and PPV$\mathrm{_{face}^{L}}$ decreases to $a$ $\approx$ 3.4, 1.3, and 0.6, respectively.
The reason for the decreasing  $a$ is that the resolution has a notable effect on the cloud surface density, where the high-${\Sigma_{\mathrm{c}}}$ population  is  missing due to the increase in radius, pushing their relation to  low-$\Sigma_\mathrm{c}$   for a given mass. 
 It is worth noting that the order of the three slopes is not changed as we progress to lower resolution.
 In PPV$\mathrm{_{face}^{L}}$,  there is an outlier group of relatively massive clouds sitting at mass  $\geq$ 10$^{6}$ M$_{\sun}$ but with radii of $\leq$ 20 pc. The radii of these massive non-star-forming clouds are likely underestimated due to the high density cores taking over the weighting. These clouds occupy about 10\% of total populations. Even  excluding these clouds arbitrarily, $M_{\mathrm{c}}$-$R_{\mathrm{c}}$ relation of  PPV$\mathrm{_{face}^{L}}$ is still flatter than that of  PPV$\mathrm{_{edge}^{L}}$.  
  
Low-resolution observations of nearby galaxies obtain an index $a$ in the range of $\sim$ 1.5 -- 2.6 \citep{Eng03,Bol08,Reb12,Don13,Col14,Uto15}. The values are larger than our derived values for both PPV$\mathrm{^{H}}$.  In observations, $M_{\mathrm{c}}$  is usually  calculated by adopting a constant $X_{\mathrm{CO}}$ for all clouds,  or  represented by the measured CO luminosity ($L_{\mathrm{CO}}$). However, there is growing evidence that  $X_{\mathrm{CO}}$ is not constant among GMCs. Small clouds which  have  higher fraction of CO-dark molecular gas 
  require  larger $X_{\mathrm{CO}}$ to  recover the  H$_{2}$ mass from  $L_{\mathrm{CO}}$.  Thus we would expect that the  observed $M_{\mathrm{c}}$-$R_{\mathrm{c}}$ of nearby galaxies would become shallower once the variable  $X_{\mathrm{CO}}$ is considered.

\subsubsection{Velocity Dispersion--Radius Relation}

Results of $\sigma_{\mathrm{c}}$-$R_{\mathrm{c}}$ are shown in the middle panels of Figure \ref{FIG_larsons}.
 The best fits of all clouds produce a gradient of $b$ $\approx$ 1.2 for PPP$\mathrm{^{H}}$ and $b$ $\approx$ 1.0 for PPP$\mathrm{^{L}}$.
Variation of $\sigma_{\mathrm{c}}$-$R_{\mathrm{c}}$ relation is seen for both resolutions  at low ($<$ 4 km s$^{-1}$) and high ($>$ 4 km s$^{-1}$)  $\sigma_{\mathrm{c}}$.
We thus also perform the fit only considering those structures separately, yielding a $b$ of 0.50 $\pm$ 0.05 and 0.46 $\pm$ 0.03 for the low-$\sigma_{\mathrm{c}}$ clouds in PPP$\mathrm{^{H}}$ and PPP$\mathrm{^{L}}$, and $b$ of 1.13 $\pm$ 0.02 and 1.05 $\pm$ 0.06 for the high-$\sigma_{\mathrm{c}}$ structures, respectively.
The results suggest that resolution has relatively small effect on both Larson's relations when the cloud properties are extracted from 3 dimensional measurements. 
 But note that the high and low-$\sigma_{\mathrm{c}}$ (or $\Sigma_{\mathrm{c}}$) clouds are  relatively discrete in the high-resolution relations, while they are  continuous in the low-resolution  relations due to the removal of the three types populations.

For the PPV clouds, there is a large discrepancy in the $\sigma_{\mathrm{c}}$-$R_{\mathrm{c}}$ relation between the four datasets.
As seen in the $M_{\mathrm{c}}$-$R_{\mathrm{c}}$ relation, the best-fitting power law indices flatten from high to low resolution, and from edge-on to face-on observation with $b$ $\approx$ 0.8 and 0.5 for  PPV$\mathrm{_{edge}^{H}}$  and PPV$\mathrm{_{face}^{H}}$ and  $b$ $\approx$ 0.4 and 0.0 for  PPV$\mathrm{_{edge}^{L}}$ and  PPV$\mathrm{_{face}^{L}}$, respectively.
 Observations also suggest a large range of index between $b$ $\approx$ 0 -- 2, and there is no obvious correlation between the observed resolution, disc inclination and the derived index \citep{Eng03,Ros03,Ros07,Bol08,Won11,Don13,Col14,Reb15,Uto15}.
Interestingly, the  weak to no correlations of PPV$\mathrm{_{edge}^{L}}$ and PPV$\mathrm{_{face}^{L}}$ are in line with the latest unresolved observations of GMCs  \citep{Col14,Reb15,Uto15,Ler15}.

In fact, it is unlikely  that the $\sigma_{\mathrm{c}}$-$R_{\mathrm{c}}$ relation would be identical between PPV datasets when both variables are projection-dependent but molecular clouds have  non-spherical shapes \citep[e.g.,][]{Fal91,She10,Kho15}.  For example, if the LOS goes along the short dimension of the cloud, the use of PPV can result in an overestimation  of $R_{\mathrm{c}}$ along with an underestimation of  $\sigma_{\mathrm{c}}$. 
This can explain why  the high resolution face-on observation suggest large amount of clouds with low $\sigma_{\mathrm{c}}$ (Figure  \ref{FIG_PDF_vel} and \ref{FIG_larsons_high}).

A notable feature of the power-law indices $a$ and $b$ is the marked difference across low-resolution datasets compared to high resolution. 
The discrepancy arises maybe because the resolution of  $\sim$ 24  pc can randomly sample various combination of cloud (or ISM) structures \citep[see also][]{Cal12}.
In other words, the resolution  resolves neither the individual GMCs nor the full cloud mass spectrum, which may be used to interpret the composition of GMCs within an area.
The measured cloud properties  are therefore sensitive to  their intrinsic properties and the geometry of cloud distribution, which includes the intrinsic and  projected distributions.
 Hence,  the derived cloud properties and the scaling relation with this resolution are not entirely  driven by the underlying physical origin.

Our results of two Larson's relations suggest that  such scaling relations may not genuinly reflect the physical properties  and the dynamical state of GMCs. 
Therefore, Larson's relations  should not be used alone to interpret the  physical properties and the environmental dependence of GMCs.

\subsubsection{Virial Parameter-Radius Relation}

\begin{figure*}
	\begin{subfigure}[]{0.48\textwidth}
		\includegraphics[width=1\textwidth]{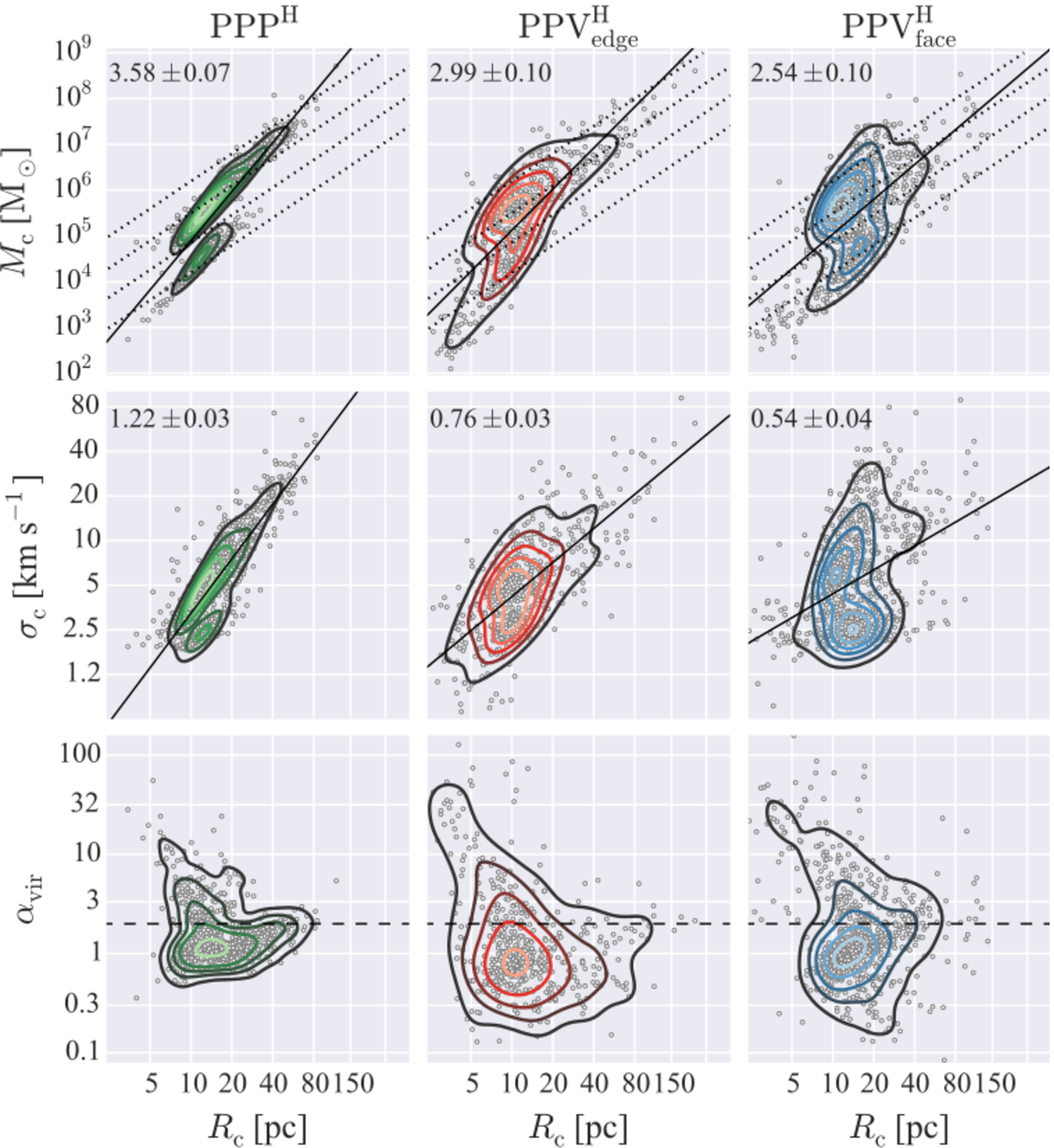}
		\caption{}\label{FIG_larsons_high}	
	\end{subfigure}
	\begin{subfigure}[]{0.48\textwidth}
		\includegraphics[width=1\textwidth]{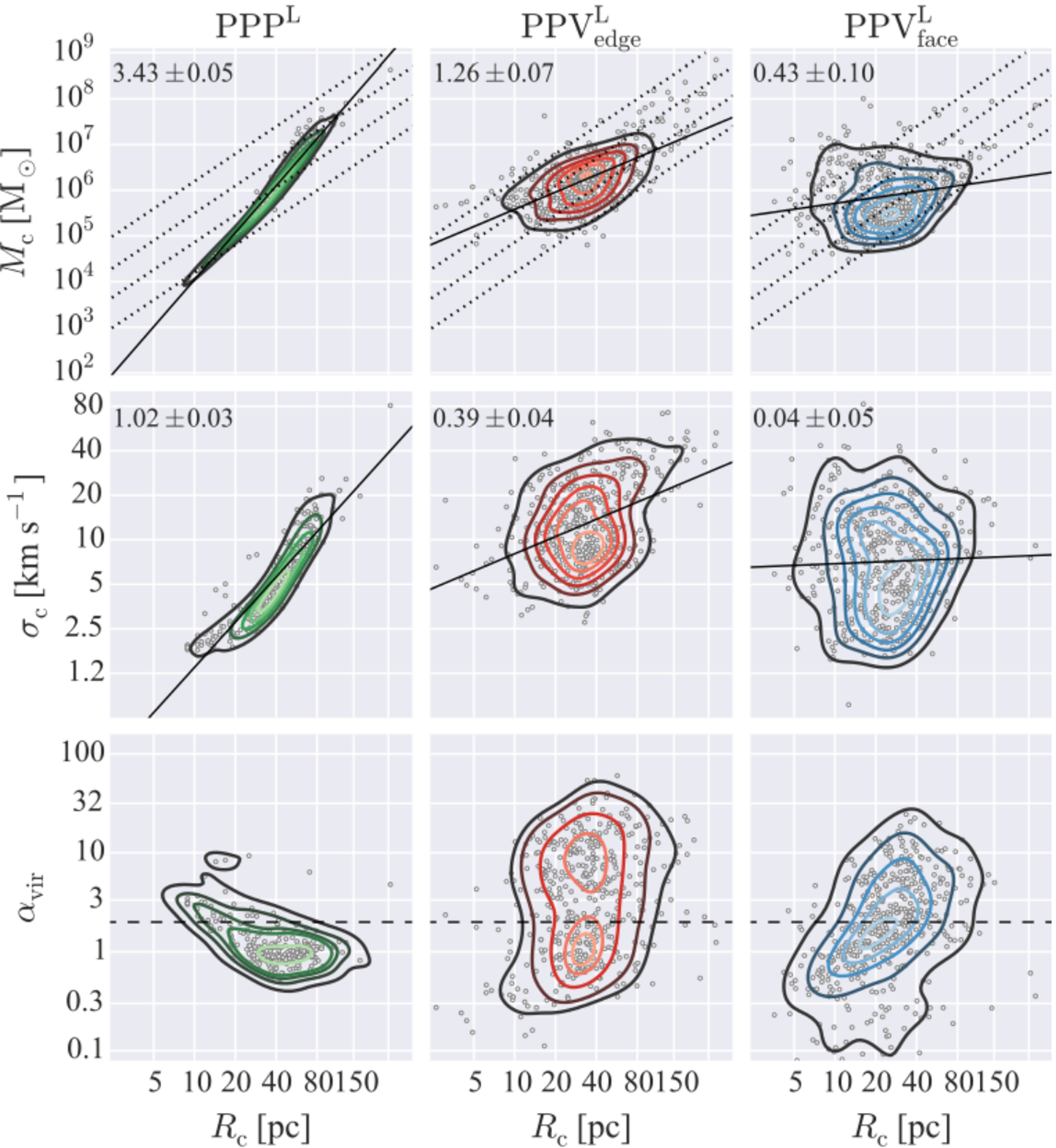} 
		\caption{}\label{FIG_larsons_low}	
	\end{subfigure}
	\caption{Larson's scaling relations for the identified clouds. From the top to bottom rows, relations for mass-radius ($M_{\mathrm{c}}$ -- $R_{\mathrm{c}}$), velocity dispersion-radius ($\sigma_{\mathrm{c}}$ -- $R_{\mathrm{c}}$), and virial parameter-radius ($\alpha_{\mathrm{c}}$ -- $R_{\mathrm{c}}$) are presented, respectively. The horizontal dashed line in the  $\alpha_{\mathrm{vir}}$ -- $R_{\mathrm{c}}$ relation marks the boundary of virial equilibrium, where $\alpha_{\mathrm{c}}$ $=$ 2. The left to right columns show the results of PPP,  PPV$_{\mathrm{edge}}$, and  PPV$_{\mathrm{face}}$, respectively. Power law index of the first two relations are given in the upper-left corner of the panels. (a) Results with high resolution.  Dotted lines  denote $\Sigma_{\mathrm{c}}$ $=$ 50 (lower), 230, 1000, and 5000 (upper) M$_{\sun}$ pc$^{-2}$, respectively. (b) Results with low resolution. It is clear that both resolution and inclination effects influence the observed GMC properties and the slopes of Larson's scaling relations.}
	\label{FIG_larsons}
\end{figure*}

PPP$\mathrm{_{edge}^{H}}$  clouds show two populations on the $\alpha_{\mathrm{vir}}$-$R_{\mathrm{c}}$ plane in the bottom row of Figure \ref{FIG_larsons_high}.
The  unbound population ($\alpha_{\mathrm{vir}}$ $>$ 2) sitting at the small $R_{\mathrm{c}}$ regime  shows decreasing $\alpha_{\mathrm{vir}}$ with  $R_{\mathrm{c}}$,  i.e., the smaller clouds are the least gravitationally bound.
These are  the `Type C' clouds with  $\Sigma_{\mathrm{c}}$ $<$ 230 M$_{\sun}$ pc$^{-2}$. 
The bound population  with $\alpha_{\mathrm{vir}}$ $\approx$ 1 spreads out over a wide range of $R_{\mathrm{c}}$ and show $\alpha_{\mathrm{vir}}$ increasing with $R_{\mathrm{c}}$.
These  are  the `Type A' and `Type B'  with high $\Sigma_{\mathrm{c}}$.

The three type clouds and their distinct $\alpha_{\mathrm{vir}}$ are  reproduced  but more scattered  in  PPV$\mathrm{_{edge}^{H}}$ and PPV$\mathrm{_{face}^{H}}$ clouds \citep{Pan15}.
Overall, PPV$\mathrm{_{face}^{H}}$ clouds share  the same features as PPP$\mathrm{^{H}}$.
For the PPV$\mathrm{_{edge}^{H}}$ clouds,  even though the  Type C clouds are not  clear  on the $M_{\mathrm{c}}$-$R_{\mathrm{c}}$ relation, it is visible in the $\alpha_{\mathrm{vir}}$-$R_{\mathrm{c}}$ relation. 
This population has considerably high $\alpha_{\mathrm{vir}}$ and show the same variation in $R_{\mathrm{c}}$  as seen in PPP$\mathrm{^{H}}$ and PPV$\mathrm{_{face}^{H}}$.
The bound population of PPV$\mathrm{_{edge}^{H}}$, however,  shows a weak correlation between $\alpha_{\mathrm{vir}}$ and $R_{\mathrm{c}}$. The values are concentrated around 1 from $R_{\mathrm{c}}$ $\approx$ 5 -- 50 pc, but slightly increase toward $R_{\mathrm{c}}$ $\approx$ 80 pc. This is because  $R_{\mathrm{c}}$ of PPV$\mathrm{_{edge}^{H}}$ clouds are relatively uniform around 10 pc (Figure \ref{FIG_PDF_rad}) due to  the flat galactic disc. 

Turning to the low resolution, PPP$\mathrm{^{L}}$  clouds show similar features to PPP$\mathrm{^{H}}$ in the $\alpha_{\mathrm{vir}}$-$R_{\mathrm{c}}$ relation, however, the variation of cloud properties  have a very notable effect on $\alpha_{\mathrm{vir}}$ for   the low-resolution PPV cloud   (bottom rows of Figure \ref{FIG_larsons_low}).
$\alpha_{\mathrm{vir}}$ for PPV$\mathrm{_{face}^{L}}$ clouds increase with $R_{\mathrm{c}}$, indicating that the large clouds are least bound.  The pattern and the underlying implication are similar to the bound population in its high-resolution counterpart, but the relation is much more scattered.
 We note that a high-$\alpha_{\mathrm{vir}}$ cloud is harder to detect than a lower-$\alpha_{\mathrm{vir}}$ cloud with the same $M_{\mathrm{c}}$ because the spread over larger $R_{\mathrm{c}}$ or $\sigma_{\mathrm{c}}$ will make the surface brightness per channel lower.
 PPV$\mathrm{_{edge}^{L}}$ clouds show different distribution characteristics, including two concentrations with a dominant population at  $\alpha_{\mathrm{vir}}$ $\approx$ 1 and a secondary  at  $\alpha_{\mathrm{vir}}$ $\approx$ 3 -- 10.  We found that the  unbound populations tend to have relatively large  $\sigma_{\mathrm{c}}$ and small $M_{\mathrm{c}}$.   Because $\alpha_{\mathrm{vir}}$  is proportional to the square of $\sigma_{\mathrm{c}}$ and inverse of $M_{\mathrm{c}}$, $\alpha_{\mathrm{vir}}$ then increases significantly.
This is in agreement with previous argument that small clouds are  more susceptible to (velocity) blending. 
The blended emission can  spread out over a wide velocity channels, but  $M_{\mathrm{c}}$ does not increase as fast as $\sigma_{\mathrm{c}}$ since they are intrinsically small.

The large discrepancy in $\alpha_{\mathrm{vir}}$ can lead to inaccurate interpretations of the dynamical state of GMCs, and therefore their potential for star formation \citep[e.g.,][]{Bal06}.
\cite{She10} and \cite{Bea13}  have also recognized the difficulty to determine $\alpha_{\mathrm{vir}}$  in their simulated PPV clumps   even though the clumps are  more compact with less substructure than our GMCs. 
\cite{She10} further suggest that the classic derivation of $\alpha_{\mathrm{vir}}$ for simple spherical structures may not be sufficient to reliably determine if a GMC  is bound or not. 
Revision is required to handle the non-spherical shapes and  projection effects and to include additional physics in both PPP and PPV clouds, such as  $\Sigma_{\mathrm{c}}$, magnetic fields, and time \citep{Ber92,Bal06,Dib07,She10,Her11}.
That is, dynamical state of the GMCs may not be single value to be determined.  

 Moreover, our results suggest that the  virial mass ($M_{\mathrm{vir}}$)-based analysis of extragalactic $X_{\mathrm{CO}}$ should be used with caution.  Many studies have used the classically derived $M_{\mathrm{vir}}$ (\S\ref{sec_def_gmc}) of GMCs  to estimate the extragalactic $X_{\mathrm{CO}}$ assuming virial equilibrium \citep[$M_{\mathrm{vir}}$ $\approx$ $M_{\mathrm{c}}$, e.g.,][]{Adl92b,Isr03,Ros07,Bol08,Hug10,Don12,Reb12}.  
  However,  our results show that  $M_{\mathrm{vir}}$ is projection  dependent in the PPV space, and therefore the derived $X_{\mathrm{CO}}$ would be affected by the observational bias as well. 
 This is particularly true for the low, but realistic, resolution. 
 The estimation of $X_{\mathrm{CO}}$ can be improved by increasing the observed resolution as suggested by the similar $\alpha_{\mathrm{vir}}$ between the high-resolution datasets (Figure \ref{FIG_PDF_alpha}), however, the required resolution can be rather unrealistic for extragalactic observations even with ALMA, e.g., the 1.5 pc in this work.  
 

The results for $\alpha_{\mathrm{vir}}$  also imply a significant role for the cloud definition method and selection criteria  \citep[e.g.,][]{Iss90,She08,She12,Hug13,Fuj14,Pan15, Col15}. There is  no obvious edge to the clouds, which are often thought to be borderline of gravitationally bound. However, the required boundary that truncates the cloud from a continuous ISM can change  if the clouds are observed from a different  direction, different resolution, and different data structure.  We emphasize that our analysis is on the basis of a specific cloud identification method which can potentially identify the same objects with close properties between simulations and observations \citep{Pan15}. It is not necessarily  the most suitable method for other datasets, depending on the data quality and the scientific goal.  The major reason for the ambiguity here is the absence of a practical ``definition'' of a giant molecular cloud.

\section{Summary}
\label{sec_summary}

While ALMA is about to resolve a wide population of GMCs across different galaxy environments, understanding the observational bias is essential  for obtaining reliable GMC properties. Observational bias, such as the disc inclination and the observed resolution in this work, is assessable by comparing GMC properties  found when using observational identification techniques to those typically used for simulation data.
 To achieve this, we compared the physical properties of GMCs formed in a simulation of a barred spiral galaxy using both simulation and observational identification methods.
The two methods identified clouds in the data using the PPP space typical for simulations and the PPV space used in observations.  In each case, two resolutions were considered: one at the maximum resolution of the simulation data (1.5\,pc) and another at a more realistic level for observational instruments (24\,pc). For the PPV data, the galaxy disc was also considered both face-on and side-on, to explore the result of projection effects.
The PPV data cube was assumed to be the product of $^{12}$CO (1--0) observations.

The main results for the high-resolution (1.5 pc) cloud properties in PPP$^{\mathrm{H}}$, PPV$\mathrm{^{H}_{edge}}$ and  PPV$\mathrm{^{H}_{face}}$ are as follows:
\begin{enumerate}

\item The galactocentric profile of cloud numbers are similar in all datasets, showing two concentrations at the radii of galactic bar and spiral arms. However, the cloud number in the PPV$\mathrm{^{H}_{edge}}$ analysis lies below the PPP$^{\mathrm{H}}$ and PPV$\mathrm{^{H}_{face}}$ datasets by roughly a factor of two, indicating that the clouds are blending when viewed edge-on in PPV space. Thus, even though the distribution and median values of the high-resolution cloud properties (mass and velocity dispersion) agree with each other,  this should not lead to the interpretation that the clouds in PPV$\mathrm{^{H}_{edge}}$ and  PPP$^{\mathrm{H}}$/PPV$\mathrm{^{H}_{face}}$ are  completely the same.
  \item Disc inclination has a notable effect on the cloud radius where the results from the PPV$\mathrm{^{H}_{edge}}$ become smaller than the other two datasets due to the flat galactic disc.
  \item The bimodal mass surface density distribution of PPP$^{\mathrm{H}}$ clouds as a result of different formation mechanisms suggested in \cite{Fuj14}, is reproduced by both the PPV$\mathrm{^{H}_{edge}}$ and PPV$\mathrm{^{H}_{face}}$ methods, although the boundary between the two trends is least distinguishable for PPV$\mathrm{^{H}_{edge}}$, due to small populations being particularly  susceptible to blending.
  \item The data structure and disc inclination do not significantly affect the virial parameter of the high resolution clouds. All methods determined a median virial parameter  of $\sim$ 1.0, suggesting that majority of clouds are gravitationaly bound.
\end{enumerate}

We prepared low-resolution (24 pc) PPV datasets using CASA, allowing us to simulate  ALMA observation with realistic resolution and sensitivity (noise level).  The low-resolution PPP clouds were identified from a simulation with fewer total levels of refinement.
The main results for the low-resolution cloud properties in the PPP$^{\mathrm{L}}$, PPV$\mathrm{^{L}_{edge}}$ and  PPV$\mathrm{^{L}_{face}}$ datasets are:
\begin{enumerate}
\item We first compared the results from PPP$^{\mathrm{L}}$ and PPP$^{\mathrm{H}}$ to evaluate the blending (or resolution) effect alone. This revealed that although the cloud properties change, the simulation can potentially see the spatial distribution and dynamical state of clouds regardless of resolution. However, it is not able to distinguish clouds formed via different formation mechanisms  at 24 pc resolution, even though the cloud properties are extracted directly from  3 dimensions.
\item When we switched from high to low resolution (and sensitivity) in observations, the smaller clouds are lost. Large clouds are detected but become more spherical than their intrinsic morphology due to the image convolution with the nearly circular beam.
  \item   The galactocentric profile of  PPV$\mathrm{^{L}_{edge}}$ no longer represents the intrinsic distribution of clouds, but declines with radius, whereas  PPV$\mathrm{^{L}_{face}}$  show a similar profile as   PPP$^{\mathrm{L}}$ and their high-resolution counterparts, but with fewer clouds at all radii.
  \item  Overall, we found good agreement between the profiles and median values of the cloud properties between the  PPV$\mathrm{^{L}_{face}}$ and PPP$\mathrm{^{L}}$ techniques, while PPV$\mathrm{^{L}_{edge}}$ clouds are more massive and with larger velocity dispersions.  
  \item    In contrast with the high-resolution data, the mass density is not bimodal in any of the low-resolution data.
  \item  The resolution has an alarming effect on the virial parameter. Median $\alpha_{\mathrm{vir}}$ of PPV$\mathrm{^{L}_{edge}}$ suggests that the clouds are not gravitationally bound, which is in contrast to  PPP$\mathrm{^{L}}$  and PPV$\mathrm{^{L}_{face}}$, where the clouds seem to be bound.
 
\end{enumerate}

We plotted  Larson's scaling relations of mass--radius ($M_{\mathrm{c}}\propto R\mathrm{_{c}}^{a}$) and velocity dispersion--radius ($\sigma_{\mathrm{c}}\propto R\mathrm{_{c}}^{b}$) using the measured cloud properties,  the main results are:

\begin{enumerate}
  \item The power-law indices of the Larson relations change with data structure, disc inclination, and resolution. In all relations, the indices decrease from PPP, PPV$_{\mathrm{edge}}$ to PPV$_{\mathrm{face}}$ and decrease from high to low resolutions. 
  \item For individual data structures, the power-law index of PPP is relatively insensitive to resolution, while it changes by $\geq$ 2 times between  the two resolutions for both face-on and edge-on PPV. This suggests that  Larson's scaling might be  reliable only for the clouds properties extracted from 3 dimensions. 
  \item A discrepancy in the power-law indices between face-on and edge-on observations are also observed. Moreover, the low, but realistic, resolution  shows larger discrepancy compared to high resolution. Perhaps because a 24 pc resolution can sample clouds with various combinations, the measured cloud properties and the scaling relations are therefore not representing the true cloud properties.

\end{enumerate}
We also compared the relation between the virial parameter ($\alpha\mathrm{_{vir}}$) and $R\mathrm{_{c}}$.  Results suggest that such scaling relations are not entirely driven by the underlying physical origin of the GMCs. Therefore they should be used with caution when discussing the environmental dependence and the dynamical state of GMCs.

We made a few comments on the  CO-to-H$_{2}$ conversion factor ($X_{\mathrm{CO}}$) in real observations. First of all, the differences in the cloud mass  between the low-resolution datasets are comparable to the uncertainty of $X_{\mathrm{CO}}$. Thus the observational bias can be obscured by adopting a different $X_{\mathrm{CO}}$.  Secondly, that a  virial mass-based derivation of extragalactic $X_{\mathrm{CO}}$ should be used with caution. 
   Our results show that  $M_{\mathrm{vir}}$ is projection  dependent in the PPV space, especially with the low but realistic resolution (24 pc). Therefore the derived $X_{\mathrm{CO}}$ would be affected by the observational bias as well. 
 The estimation of $X_{\mathrm{CO}}$ can be improved by increasing the observed resolution. However, the required resolution can be rather unrealistic for extragalactic observations even with ALMA, e.g., the 1.5 pc in this work.

Finally we note that this work is based on a specific setup for a simulated galaxy. Observational bias  would  alter the GMC properties and Larson's relations in different ways if the galaxy type is changed, such as the global morphology and gas content,  because they determine the intrinsic distribution and properties of the GMCs.

\section*{ACKNOWLEDGMENTS}
We thank referee for providing useful comments that have helped to improve the paper. 
The authors would like to thank the yt development team \citep{Tur11} for support during the analysis of these simulations. Numerical computations were carried out on the Cray XT4 and Cray XC30 at the Center for Computational Astrophysics (CfCA) of the National Astronomical Observatory of Japan (NAOJ). EJT is funded by the MEXT grant for the Tenure Track System. SMB acknowledges financial support from the Vanier Canada Graduate Scholarship program.

\end{document}